\UseRawInputEncoding
\documentclass[11pt,a4paper,useAMS,usenatbib]{emulateapj}
\usepackage[T1]{fontenc}
\usepackage{epsfig}
\usepackage{amsmath}
\usepackage{natbib}

\begin{document}

\title{Exploring Gravitationally-Lensed $z\gtrsim6$ X-ray AGN Behind the RELICS clusters}

\author{\'Akos Bogd\'an\altaffilmark{1}, Orsolya E. Kov\'acs\altaffilmark{2,1},  Christine Jones\altaffilmark{1}, \\ William R. Forman\altaffilmark{1}, Ralph P. Kraft\altaffilmark{1}, Victoria Strait\altaffilmark{3}, Dan Coe\altaffilmark{4}, and Maru{\v{s}}a Brada{\v{c}}\altaffilmark{3} }
\affil{\altaffilmark{1}Harvard Smithsonian Center for Astrophysics, 60 Garden Street, Cambridge, MA 02138, USA; abogdan@cfa.harvard.edu}
\affil{\altaffilmark{2}Department of Theoretical Physics and Astrophysics, Faculty of Science, Masaryk University, Kotl\'a\v{r}sk\'a 2, Brno, 611 37, Czech Republic}
\affil{\altaffilmark{3}Department of Physics, University of California, Davis, CA 95616, USA}
\affil{\altaffilmark{4}Space Telescope Science Institute, 3700 San Martin Drive, Baltimore, MD 21218, USA}
\shorttitle{LENSED HIGH-Z AGN BEHIND THE RELICS CLUSTERS}
\shortauthors{BOGD\'AN ET AL.}

\begin{abstract}
Although observations of high-redshift quasars demonstrate that many supermassive black holes (BHs) reached large masses within one billion years after the Big Bang, the origin of the first BHs is still a mystery. A promising way to constrain the origin of the first BHs is to explore the average properties of $z\gtrsim6$ BHs. However, typical BHs remain hidden from X-ray surveys, which is due to their relatively faint nature and the limited sensitivity of X-ray telescopes. Gravitational lensing provides an attractive way to study this unique galaxy population as it magnifies the faint light from these high-redshift galaxies. Here, we study the X-ray emission originating from 155 gravitationally-lensed $z\gtrsim6$ galaxies that were detected in the RELICS survey. We utilize \textit{Chandra} X-ray observations to search for AGN in the individual galaxies and in the stacked galaxy samples. We did not identify an individual X-ray source that was undoubtedly associated with a high-redshift galaxy.  We stack the signal from all galaxies and do not find a statistically significant detection. We split our sample based on stellar mass, star-formation rate, and lensing magnification and stack these sub-samples. We obtain a $2.2\sigma$ detection for massive galaxies with an X-ray luminosity of $(3.7\pm1.6)\times10^{42} \ \rm{erg \ s^{-1}}$, which corresponds to a $(3.0\pm1.3)\times10^5 \ \rm{M_{\odot}}$ BH accreting at its Eddington rate. Other stacks remain undetected and we place upper limits on the AGN emission. These limits imply that the bulk of BHs at $z\gtrsim6$ either accrete at a few percent of their Eddington rate and/or are $1-2$ orders of magnitude less massive than expected based on the stellar mass of their host galaxy.  
\end{abstract}

\keywords{galaxies: active --- galaxies: high-redshift --- galaxies: quasars: supermassive black holes ---  X-rays: galaxies --- X-rays: general --- X-rays: galaxies: clusters}

\section{Introduction}
\label{sec:intro}

In the past decades, deep surveys detected more than 200 optically bright quasars at high ($z\gtrsim 6$) redshifts. These discoveries demonstrate that accretion-powered BHs with masses of $\sim$$10^9 \ \rm{M_{\odot}}$ were in place merely one billion years after the Big Bang \citep[e.g.][]{fan06,willott07,jiang08,mortlock11,venemans13,banados14,wang17,wang19,yang17,yang19,yang20}. The origin and rapid assembly of BHs can be explained by a number of seeding models, which are usually grouped into ``\textit{light seed}'' and ``\textit{heavy seed}'' models. The light seed scenario involves the collapse of Population III stars leading to BH seeds with $10-100 \ \rm{M_{\odot}}$ \citep[e.g.][]{madau01,volonteri03}. These low-mass seeds must grow rapidly via accretion and/or mergers to increase their mass by many orders of magnitude in less than one billion years. The heavy seed scenario involves the formation of massive, $10^4-10^5 \ \rm{M_{\odot}}$, BHs through the direct collapse of massive gas clouds \citep[e.g.][]{volonteri05,begelman06,lodato06,wise19}. Due to their large initial mass, BHs may grow through episodic accretion, hence this is an attractive model to explain the existence of luminous quasars at $z\gtrsim6$.

Understanding the origin and early growth of BHs is arguably one of the most thrilling quests of modern astrophysics. However, to constrain the formation scenarios and early growth of BHs, it is essential to probe BHs residing at high-redshift. Indeed, only these objects can provide the much-needed observational constraints. However, observations of these distant BHs are exceptionally demanding with present-day X-ray observatories. The main difficulty  in detecting the ``average'' high-redshift accreting BHs is due to their low luminosities and the relatively low sensitivity of present-generation X-ray telescopes. Despite the challenging nature, several studies attempted to detect high-redshift BHs in the X-ray waveband. X-ray follow-up of optically-identified quasars led to the detection of high-redshift AGN. Specifically, \textit{Chandra} detected several AGN at  $z\sim6$  \citep[e.g.][]{ai16,gallerani17,nanni17,vito19,connor20,pons20,wang21}. The masses of these X-ray-detected BHs is $\sim10^9 \ \rm{M_{\odot}}$, hence they do not represent the average AGN, which are likely several orders of magnitude less massive. However, these relatively low mass and high-redshift accreting BHs remained hidden from X-ray observations, and, hence our understanding of the average properties of $z\sim6$ AGN is still lacking despite the substantial efforts over the past decade. 

To characterize the average properties of AGN at medium-redshift ($z=2-5$) and high-redshift ($z=5-6$), most studies focused on the Chandra Deep Field South  \citep[CDF-S; e.g.][]{giacconi01,luo17}. By stacking the X-ray photons of a large number of galaxies, \citet{vito16} detected signals from accreting BHs at $z\approx 4$ and $z\approx 5$, but $z\approx 6$ galaxies remained undetected. Recently, \citet{liu21} utilized the Cluster Lensing And Supernova survey with Hubble (CLASH) clusters \citep{postman12} to search for AGN in medium- to high-redshift galaxies. They detected a handful of individual AGN in the redshift range of $z=2.8-5$ behind the CLASH clusters and they also stacked their galaxy sample, which led to detections in these redshifts. While these authors demonstrated the feasibility and the powerful nature of using lensing clusters to study high-redshift AGN, their study did not constrain AGN at $z\gtrsim6$.  

\begin{table*}
\caption{Characteristics of the galaxy clusters analyzed in this paper}
\begin{minipage}{18cm}
\renewcommand{\arraystretch}{1.3}
\centering
\begin{tabular}{cccccccc}
\hline
	Cluster name & RA & Dec & Redshift & Planck mass & $N_{\rm gal}$  & $t_{\rm{exp}}$  \\ 
	& (J2000)& (J2000) & & ($10^{14} \ \rm{M_{\odot}}$) & & (ks) \\ 
	(1) & (2) & (3) & (4) & (5) & (6) & (7) \\ \hline
	Abell 1300 & 11:31:54.1 & $-$19:55:23.4 & 0.307 & 8.97 & 2 & 93.5 \\ 
        Abell 1758 & 13:32:39.0 & +50:33:41.8 & 0.280 & 8.22 & 5 & 213.0 \\ 
        Abell 1763 & 13:35:18.9 & +40:59:57.2 & 0.228 & 8.13 & 8 & 94.5 \\ 
        Abell 2163 & 16:15:48.3 &  $-$06:07:36.7 & 0.203 &16.12& 5 & 80.6 \\ 
        Abell 2537 & 23:08:22.2 &  $-$02:11:32.4 & 0.297 & 5.52 & 2 & 74.7 \\ 
        Abell 2813 & 00:43:25.1 &  $-$20:37:14.8 & 0.292 & 8.13 & 1 & 400.3 \\ 
        Abell 3192 & 03:58:53.1 & $-$29:55:44.8 & 0.425 & 7.20 & 3 & 59.3 \\ 
        Abell 520   & 04:54:19.0 & +02:56:49.0 & 0.203 & 7.80 & 4 & 528.2  \\ 
        Abell 665   & 08:30:57.4 & +65:50:31.0 & 0.182 & 8.86 & 3 & 139.6  \\
        Abell 697   & 08:42:58.9 & +36:21:51.1 & 0.282 & 11.0 & 4 & 27.5 \\ 
        Abell S295 & 02:45:31.4 & $-$53:02:24.9  & 0.300 & 6.78 & 2 & 205.7 \\ 
        CL0152-13 & 01:52:42.9 & $-$13:57:31.0 & 0.833 & $-$ & 18  & 95.3  \\ 
        MACS0159-08 & 01:59:49.4 & $-$08:50:00.0 & 0.405 & 7.20 &  3 & 72.7  \\ 
        MACS0025-12 & 00:25:30.3 & $-$12:22:48.1 & 0.586 & $-$ & 3 & 157.6 \\ 
        RXC0911+17 & 09:11:11.4 & +17:46:33.5 & 0.505 & 6.99 & 3 & 41.7  \\
        MACS0035-20 & 00:35:27.0 & $-$20:15:40.3 & 0.352 & 7.01 & 3 & 21.4 \\ 
        MACS0257-23 & 02:57:10.2 & $-$23:26:11.8 & 0.505 & 6.22 & 8 & 38.3 \\ 
        MACS0308+26 & 03:08:55.7 & +26:45:36.8 & 0.356 &10.76 & 4 & 24.4 \\ 
        MACS0553-33 & 05:53:23.1 & $-$33:42:29.9 & 0.430 & 8.77 & 6 & 84.0 \\ 
        RXC0949+17 & 09:49:50.9 & +17:07:15.3 & 0.383 & 8.24 & 3 & 14.3  \\ 
        RXC2211-03 & 22:11:45.9 & $-$03:49:44.7 & 0.397 & 10.5 & 2 & 17.7 \\ 
        RXC0018+16 & 00:18:32.6 & +16:26:08.4 & 0.546 & 9.79 & 6 & 67.4  \\ 
        MS1008-12 & 10:10:33.6 & $-$12:39:43.0 & 0.306 & 4.94 & 8 & 51.2  \\ 
        PLCKG171-40 & 03:12:56.9 & +08:22:19.2 & 0.270 & 10.71  & 2 & 26.7 \\ 
        SMACS0723-73 & 07:23:19.5 & $-$73:27:15.6 & 0.390 & 8.39  & 6 & 19.8 \\ 
        SPT0615-57 & 06:15:54.2 & $-$57:46:57.9 & 0.972 & 6.77 & 11 & 241.3 \\ 
        PLCKG287+32 & 11:50:50.8 & $-$28:04:52.2 & 0.39 &14.69  & 15 & 196.4 \\ 
        PLCKG138-10 & 02:27:06.6 & +49:00:29.9 & 0.702 & 9.48 & 3 & 11.9   \\ 
        PLCKG209+10 & 07:22:23.0 & +07:24:30.0 & 0.677 & 10.73 & 5 & 10.0   \\ 
        RXC1514-15 & 15:15:00.7 & $-$15:22:46.7 & 0.223 & 8.86  & 1 & 59.2   \\ 
        RXC0232-44 & 02:32:18.1 & $-$44:20:44.9 & 0.284 & 7.54 & 2 & 23.4  \\ 
        RXS0603+42 & 06:03:12.2 & +42:15:24.7 & 0.228 & 10.76 & 8 & 237.1 \\ 
        RXC0142+44 & 01:42:55.2 & +44:38:04.3 & 0.341 & 9.02 & 4 & 6.0 \\ 
	PLCKG004-19 & 19:17:04.50 & $-$33:31:28.5 & 0.520 & 10.36 & 14 & 96.1  \\
	
\hline
\end{tabular} 
\vspace{0.1in}
\end{minipage}
Columns are as follows. (1) Name of the RELICS cluster; (2) and (3) right ascension and declination of the galaxy cluster; (4) Redshift of the galaxy cluster; (5) Total mass of the galaxy cluster derived from Planck measurements; (6) Number of $z>5.5$ galaxies behind the galaxy cluster; (7) Total \textit{Chandra} exposure time of the galaxy cluster. 
\label{tab:clusters}
\end{table*}

\begin{figure*}[!]
  \begin{center}
    \leavevmode
      \epsfxsize=0.45\textwidth \epsfbox{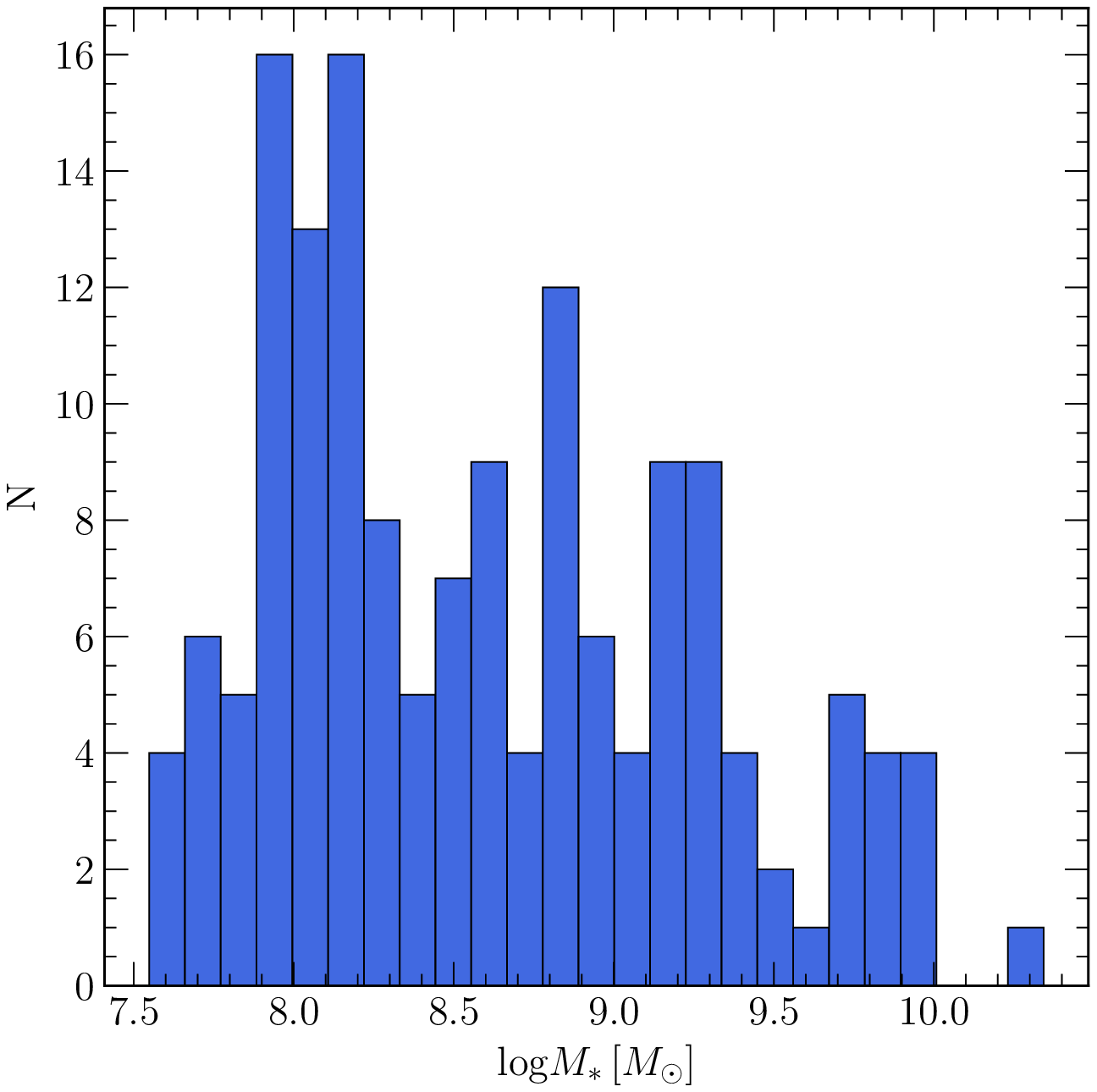}
      \epsfxsize=0.45\textwidth \epsfbox{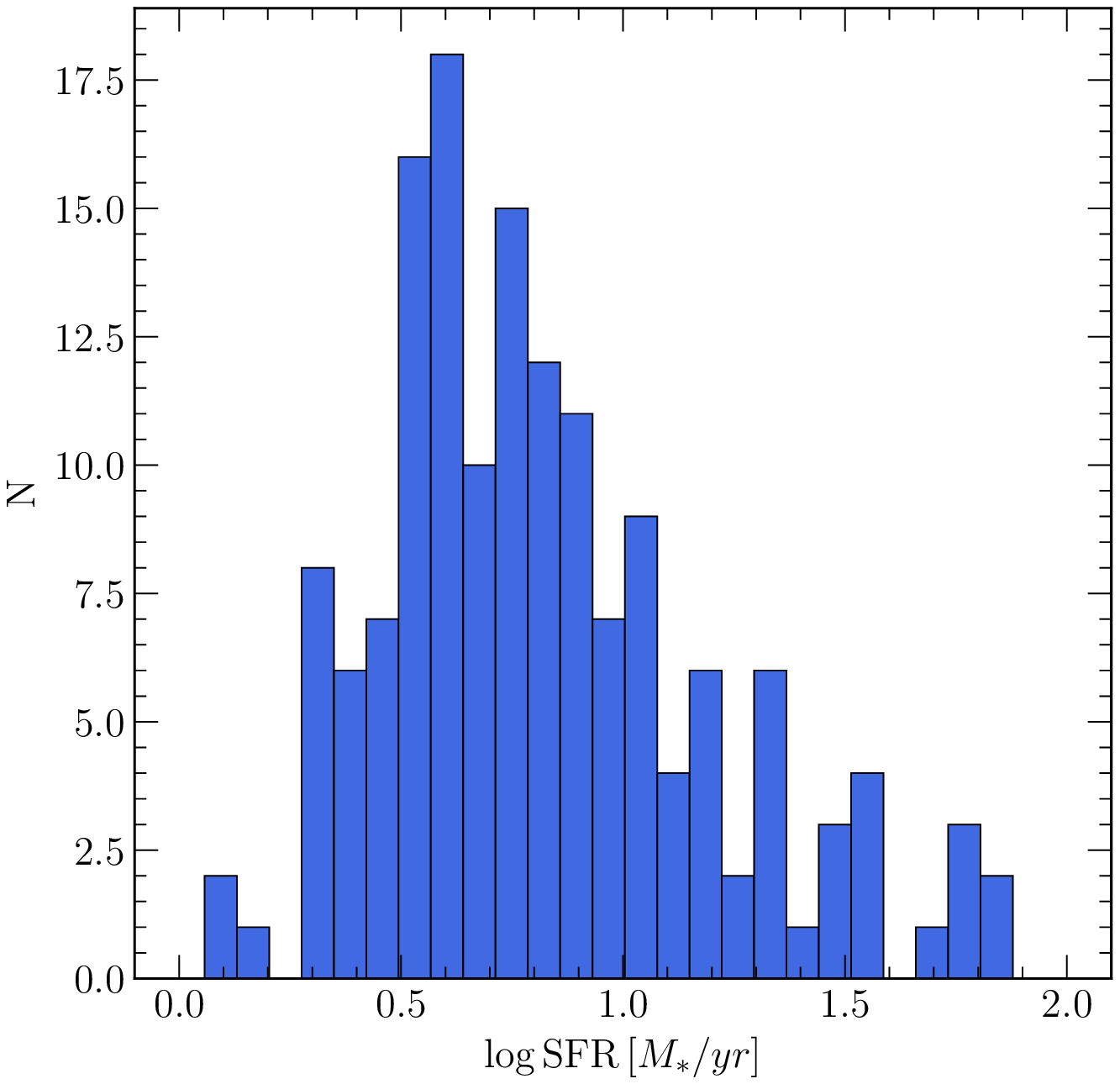}
      \epsfxsize=0.45\textwidth \epsfbox{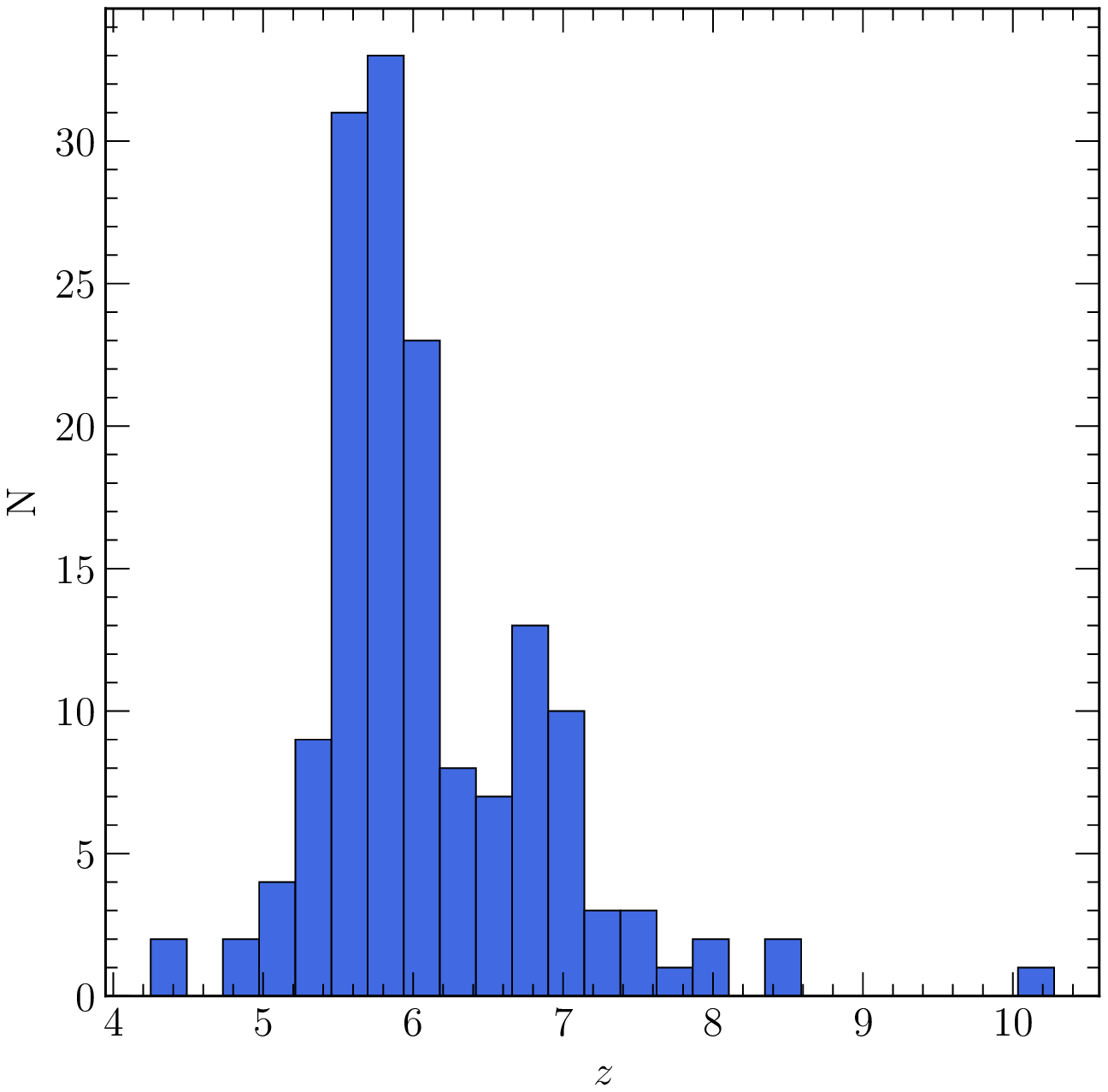}
      \epsfxsize=0.45\textwidth \epsfbox{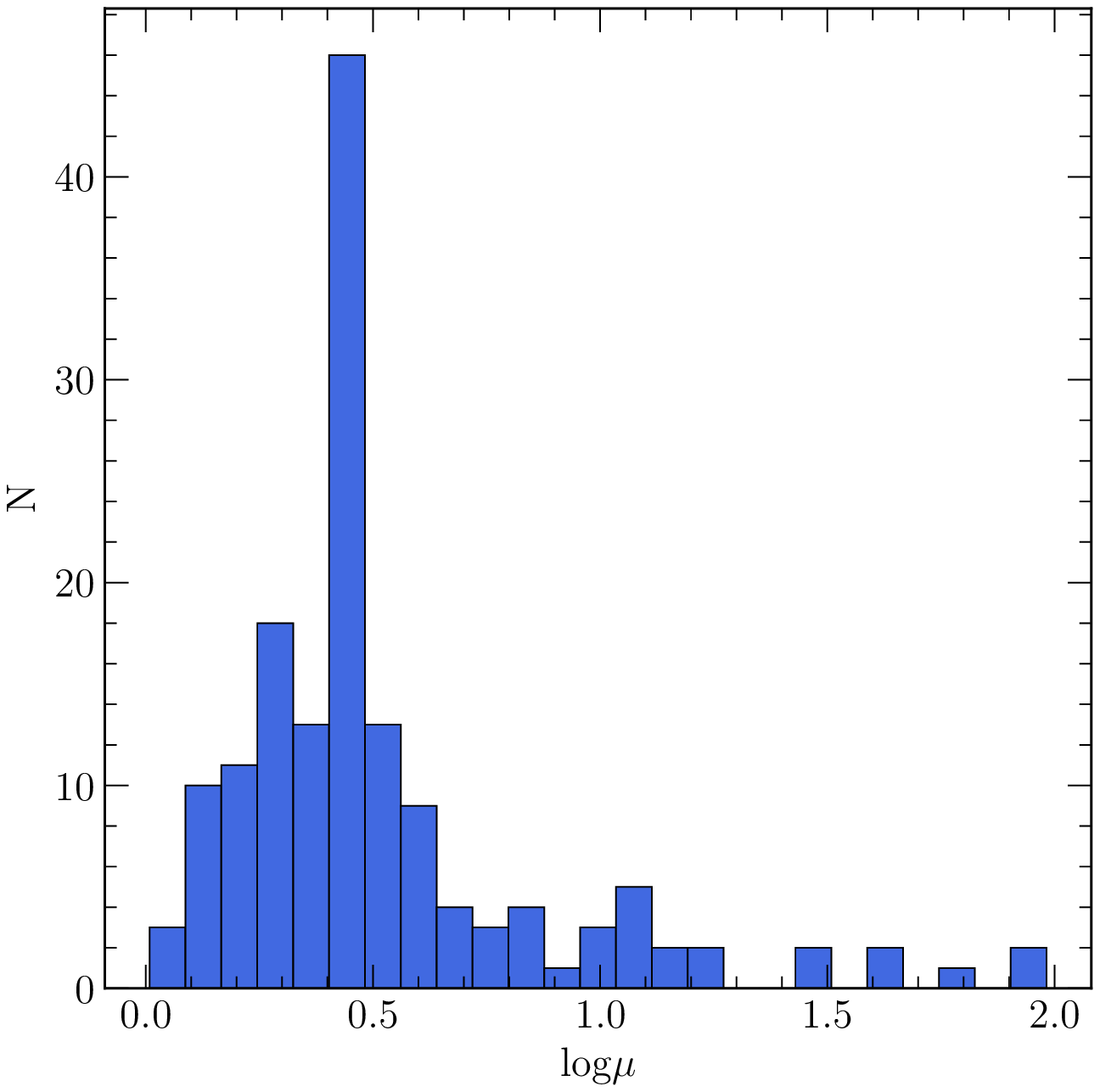}
      \vspace{0cm}
      \caption{Properties of the 155 galaxies analyzed in this work: stellar mass distribution (top left), star formation rate distribution (top right), redshift distribution (bottom left), and median lensing magnification distribution (bottom right). The values are taken from the analysis of \citet{strait20} and we relied on their Method A median values to describe the galaxy properties. The lensing magnifications are the median values obtained from different lens models (Section \ref{sec:relics}).} 
     \label{fig:distributions}
  \end{center}
\end{figure*}

In this work, we  utilize gravitational lensing that brings into focus fainter sources by magnifying them. Specifically, we rely on the rich X-ray data from the \textit{Chandra X-ray Observatory}, optical data from the \textit{Hubble Space Telescope} (HST), and infrared data obtained from the \textit{Spitzer Space Telescope} for galaxies identified by the Reionization Lensing Cluster Survey \citep[RELICS;][]{salmon20}. Through gravitational lensing, these massive galaxy clusters magnify the faint light from high-redshift galaxies behind them. Due to gravitational lensing, only the X-ray photons associated with the source are magnified, while the sky and instrumental background components and -- most importantly -- the cluster emission is not enhanced. Despite the enhanced signal, the bulk of individual AGN may be too faint to be detected individually. To this end, we boost the signal-to-noise ratios by co-adding (i.e.\ stacking) the X-ray photons from the individual galaxies. Due to the stacking approach, the combined exposure time will be increased,  which allows us to probe the X-ray flux of high-redshift accreting BHs to very low limits. Therefore, this technique could reveal a stacked detection if the individual AGN remain hidden. 

This paper is structured as follows. In Section \ref{sec:relics} we introduce the analyzed high-redshift galaxy sample and describe its properties. The analysis of the \textit{Chandra} data is described in Section \ref{sec:chandra}. The results of our paper, including the potential individual detections and the detection and upper limits on the stacked samples are presented in Section \ref{sec:results}. In Section \ref{sec:discussion}, we place our results into context, where we compare our results with previous studies, and discuss the overall importance of our results. We summarize the results of our study in Section \ref{sec:conclusions}. Throughout the paper we assume $H_0=70 \ \rm{km \ s^{-1} \ Mpc^{-1}}$, $ \Omega_M=0.27$, and $\Omega_{\Lambda}=0.73$. The error bars are $1\sigma$ uncertainties and the presented upper limits are also $1\sigma$ limits.

\begin{table*}[!t]
\begin{center}
\caption{List of analyzed \textit{Chandra} observations}
\begin{minipage}{18cm}
\renewcommand{\arraystretch}{1.3}
\centering
\begin{tabular}{ cccc|cccc}
\hline
Obs.\,ID & $t_{\rm exp} (ks)$ & Detector & Obs.\,Date & Obs.\,ID & $t_{\rm exp} (ks)$ & Detector & Obs.\,Date \\
\hline
 520    &    67.41    &    ACIS-I    &    2000-08-18     &    14349    &    24.75    &    ACIS-I    &    2012-11-09    \\  
528    &    9.47     &    ACIS-I    &    2000-10-10     &    14350    &    11.93    &    ACIS-I    &    2012-11-21    \\  
531    &    9.01     &    ACIS-I    &    1999-12-29     &    14351    &    26.23    &    ACIS-I    &    2012-11-12    \\ 
532    &    7.97     &    ACIS-I    &    1999-10-21      &   14437    &    25.94    &    ACIS-I    &    2012-09-16    \\ 
545    &    9.45     &    ACIS-I    &    2000-07-29      &   15093    &    111.22    &    ACIS-S    &    2013-09-12   \\
913    &    36.48    &    ACIS-I    &    2000-09-08      &   15094    &    141.20    &    ACIS-S    &    2013-10-24   \\
926    &    44.23    &    ACIS-I    &    2000-06-11      &   15095    &    19.81    &    ACIS-S    &    2013-09-08    \\
1653    &    71.15    &    ACIS-I    &    2001-06-16     &   15171    &    76.57    &    ACIS-I    &    2013-11-25    \\
1654    &    19.85    &    ACIS-I    &    2000-10-03     &   15172    &    111.10    &    ACIS-I    &    2013-11-28   \\
2213    &    58.31    &    ACIS-S    &    2001-08-28     &   15175    &    59.24    &    ACIS-I    &    2013-04-09    \\
3251    &    19.33    &    ACIS-I    &    2002-11-11     &   15296    &    19.82    &    ACIS-I    &    2014-04-14    \\
3262    &    21.36    &    ACIS-I    &    2003-01-22     &   15300    &    9.95     &    ACIS-I    &    2013-01-28    \\
3265    &    17.90    &    ACIS-I    &    2002-10-02     &   15302    &    26.73    &    ACIS-I    &    2013-09-26    \\
3268    &    24.45    &    ACIS-I    &    2002-03-10     &   15323    &    49.43    &    ACIS-I    &    2013-12-01    \\
3274    &    14.32    &    ACIS-I    &    2002-11-06     &   15538    &    93.33    &    ACIS-I    &    2012-09-28    \\
3276    &    13.91    &    ACIS-I    &    2002-06-14     &   15540    &    26.72    &    ACIS-I    &    2012-10-09    \\
3284    &    17.74    &    ACIS-I    &    2002-10-08     &   15572    &    14.86    &    ACIS-I    &    2012-10-29    \\
3581    &    18.47    &    ACIS-I    &    2003-08-23     &   15574    &    13.07    &    ACIS-I    &    2012-10-31    \\
3586    &    29.72    &    ACIS-I    &    2002-12-28     &   15579    &    19.82    &    ACIS-I    &    2012-11-11    \\
3587    &    17.88    &    ACIS-I    &    2003-02-23     &   15582    &    18.34    &    ACIS-I    &    2012-11-17    \\
3591    &    19.60    &    ACIS-I    &    2003-08-28     &   15588    &    23.77    &    ACIS-I    &    2012-11-22    \\
4215    &    66.27    &    ACIS-I    &    2003-12-04     &   15589    &    11.93    &    ACIS-I    &    2012-11-24    \\
4217    &    19.52    &    ACIS-I    &    2002-12-15     &   16127    &    43.32    &    ACIS-I    &    2014-07-25    \\
4962    &    36.19    &    ACIS-S    &    2004-09-09     &   16278    &    8.76     &    ACIS-I    &    2014-09-17    \\
4993    &    23.40    &    ACIS-I    &    2004-06-08     &   16282    &    9.03     &    ACIS-I    &    2014-06-13    \\
5010    &    24.83    &    ACIS-I    &    2004-08-09     &   16366    &    35.60    &    ACIS-S    &    2013-11-24    \\
5012    &    23.79    &    ACIS-I    &    2004-03-08     &   16491    &    34.07    &    ACIS-S    &    2013-11-19    \\
5813    &    9.94     &    ACIS-I    &    2005-01-08     &   16513    &    29.68    &    ACIS-S    &    2013-11-17    \\
6106    &    35.30    &    ACIS-I    &    2004-12-04     &   16524    &    44.60    &    ACIS-I    &    2014-05-20    \\
7700    &    5.08     &    ACIS-I    &    2006-12-30     &   16525    &    44.48    &    ACIS-I    &    2014-05-17    \\
7703    &    5.08     &    ACIS-I    &    2007-01-01     &   16526    &    44.48    &    ACIS-I    &    2014-08-13    \\
7710    &    6.97     &    ACIS-I    &    2007-07-12     &   17165    &    55.35    &    ACIS-I    &    2015-11-17    \\
7711    &    6.96     &    ACIS-I    &    2007-01-13     &   17166    &    20.84    &    ACIS-I    &    2015-11-24    \\
9372    &    38.51    &    ACIS-I    &    2008-08-11     &   17494    &    59.29    &    ACIS-I    &    2015-08-17    \\
9376    &    19.51    &    ACIS-I    &    2008-10-03     &   17495    &    32.23    &    ACIS-I    &    2016-03-22    \\
9409    &    19.91    &    ACIS-I    &    2008-02-02     &   18287    &    11.88    &    ACIS-I    &    2016-07-19    \\
9424    &    109.66   &    ACIS-I    &    2008-01-01     &   18292    &    9.96     &    ACIS-I    &    2015-12-07    \\
9425    &    113.52   &    ACIS-I    &    2007-12-24     &   18466    &    5.99     &    ACIS-I    &    2016-10-28    \\
9426    &    110.69   &    ACIS-I    &    2008-01-09     &   18807    &    28.70    &    ACIS-I    &    2016-03-23    \\
9430    &    113.52   &    ACIS-I    &    2008-01-11     &   19775    &    15.18    &    ACIS-I    &    2018-03-06    \\
10413   &    75.64    &    ACIS-I    &    2008-10-16     &   20590    &    19.69    &    ACIS-I    &    2019-03-03    \\
10786   &    13.91    &    ACIS-I    &    2008-10-18     &   20988   &    19.81    &    ACIS-I    &    2018-03-08    \\
10797   &    23.85    &    ACIS-I    &    2008-10-21     &    20989  &    19.79    &    ACIS-I    &    2018-03-08    \\
11719   &    9.65     &    ACIS-I    &    2009-10-18     &    20990  &    14.88    &    ACIS-I    &    2018-03-10    \\
12244   &    74.06    &    ACIS-I    &    2011-06-23     &    21702  &    35.60    &    ACIS-I    &    2019-07-30  \\
12260   &    19.79    &    ACIS-I    &    2012-01-06     &   21703  &    39.56    &    ACIS-I    &    2019-09-25  \\
12286   &    47.10    &    ACIS-I    &    2012-01-10     &   22130  &    22.76    &    ACIS-I    &    2019-03-16  \\    
12300   &    29.66    &    ACIS-I    &    2010-11-26     &   22136  &    12.79    &    ACIS-I    &    2019-02-24  \\ 
13194   &    19.97    &    ACIS-I    &    2010-11-28     &   22137  &    19.70    &    ACIS-I    &    2019-03-20  \\ 
13201   &    48.71    &    ACIS-I    &    2011-01-06     &   22679   &    26.71    &    ACIS-I    &    2019-08-02    \\
13997   &    27.64    &    ACIS-I    &    2012-09-27     &   22680   &    33.74    &    ACIS-I    &    2019-08-04    \\ 
14017   &    15.01    &    ACIS-I    &    2012-11-03     &   22856   &    19.28   &    ACIS-I    &    2019-09-26    \\
14018   &    35.60    &    ACIS-I    &    2012-09-15      \\
 \hline
\end{tabular} 
\end{minipage}
\end{center}
\label{tab:chan}
\end{table*}

\section{The analyzed galaxy sample}
\label{sec:relics}

The lensing magnification caused by the RELICS clusters provides an incredible opportunity to study high-redshift AGN behind these galaxy clusters. The RELICS survey is a \textit{Hubble Space Telescope} Treasury program (PI: Coe) and a \textit{Spitzer Space Telescope} program (PI: Brada{\v{c}}), which studied 41 massive galaxy clusters that serve as exceptional gravitational lenses. The galaxy clusters reside in the redshift range of $z=0.18-0.97$, which is ideal to utilize the lensing magnification from the clusters and study high-redshift AGN.  To study the X-ray emission from the lensed high-redshift AGN, high-resolution X-ray observations with \textit{Chandra} are essential. Although \textit{Chandra} did not carry out a systematic study of the RELICS clusters, 35 of the 41 clusters have been observed with \textit{Chandra}. Of these galaxy clusters, we excluded El Gordo (ACT-CLJ0102-49151) whose intracluster medium (ICM) is extremely bright and would outshine possible detections of faint high-redshift AGN and would dominate the overall emission in the galaxy stacks. Therefore, in this work, we study lensed galaxies that are behind the remaining 34 galaxy clusters. The basic properties of this sample are in Table \ref{tab:clusters}. 

The detailed analysis of HST and Spitzer data identified 207 galaxies with redshifts of $6 \lesssim z \lesssim8$ \citep{strait20}. Of these galaxies, 174 are behind the studied 34 galaxy clusters. Based on the \textit{HST} and \textit{Spitzer} data, the physical characteristics of the galaxies were computed using two spectral energy distribution (SED) fitting methods. In the first method (dubbed as Method A), the redshift of each galaxy was computed using the  Easy and Accurate Redshifts from Yale code \citep[EAZY;][]{brammer08}. The resulting photometric redshift probability distribution functions and stellar population synthesis templates were used to calculate the stellar properties of the galaxies. In the second method (dubbed as Method B), the Bayesian Analysis of Galaxies for Physical Inference and Parameter EStimation code \citep[BAGPIPES;][]{carnall18} was used to fit the redshift and the physical properties of the galaxies using the MultiNest nested sampling algorithm. While both of these methods utilize the SED of galaxies, they apply different methodology and template set  to obtain the galaxy properties. These differences result in different galaxy properties and distributions of these properties. \citep[for details see][]{strait20}. Following \citet{strait20}, we use the values obtained via Method A as our default throughout this paper. We note that using the redshift fitting procedure of Method A, 19 galaxies were demoted to low redshift systems. We excluded these galaxies from our study. Therefore, our final galaxy sample consists of 155 $z\gtrsim6$ galaxies. In Figure \ref{fig:distributions}, we present the stellar mass, star-formation rate, redshift, and lensing magnification distributions of the 155 lensed galaxies in our sample.

The lensing magnification of the galaxies plays an essential role in our study. In this work, we rely on the magnifications published in  \citet{strait20}. To compute the magnifications, they used three different lens models (if available) to derive lensing maps of the clusters, thereby deriving the lensing magnification for each galaxy. Since the individual lensing magnifications vary, they applied bootstrapping for each lensing model and created multiple lensing constraints for the position of each galaxy. Then the median of these realizations was computed. Since multiple methods were used to obtain the lensing maps, the final lensing magnification was the average of the medians obtained from the different methods. Further details about the analysis are provided in \citet{strait20}. The derived lensing magnifications are in the range of $\mu = 1-95$ with only 10 galaxies having $\mu > 15$. The distribution of magnifications is shown in Figure \ref{fig:distributions}. As a caveat, we note that  magnifications $\gtrsim10$ typically have large uncertainties \citep{meneghetti17}, therefore such sources need to be treated with caution.

\section{Analysis of the \textit{Chandra} data}
\label{sec:chandra}

The analysis of the \textit{Chandra} data was carried out using standard CIAO tools (version 4.13) and the CALDB version 4.9.4. All \textit{Chandra} data were obtained from the public archive. We analyzed 105 high-resolution imaging observations (Table 2). While most of these were taken with ACIS-I array, 8 observations were done with the ACIS-S array. To maximize the signal-to-noise ratios of our analysis, we include data from both ACIS-I and ACIS-S observations. The total exposure time of the observations is $3.53$~Ms. 

The first step of the analysis was to reprocess all observations. Since the observations were taken in a broad timeframe (from 2000 to 2019), it is essential to apply the same calibration data for each observation. Therefore, we used the \textit{chandra\_repro} tool on all observations. Following this step, we filtered the high background time periods. We produced light curves in the $2.3-7.3$~keV energy range in $200$~s bins and excluded those time periods where the count rate exceeded the mean value by $2\sigma$. Because ACIS-I is only weakly sensitive to high background periods, the total exposure times were not affected significantly. 

To account for vignetting effects, we constructed exposure maps for each observations. Because the main goal of this work is to study the characteristics of high-redshift AGN, we assumed a  power law model with a slope of $\Gamma=1.9$, which is appropriate to describe the spectrum of high-redshift AGN \citep{nanni17}. These exposure maps were used to convert the counts to flux units. Since several clusters are observed in multiple pointings, the individual observations and exposure maps were co-added to obtain merged event files and images of the clusters.

\begin{figure*}[!t]
  \begin{center}
    \leavevmode
      \vspace{0.25cm}
      \epsfxsize=0.85\textwidth \epsfbox{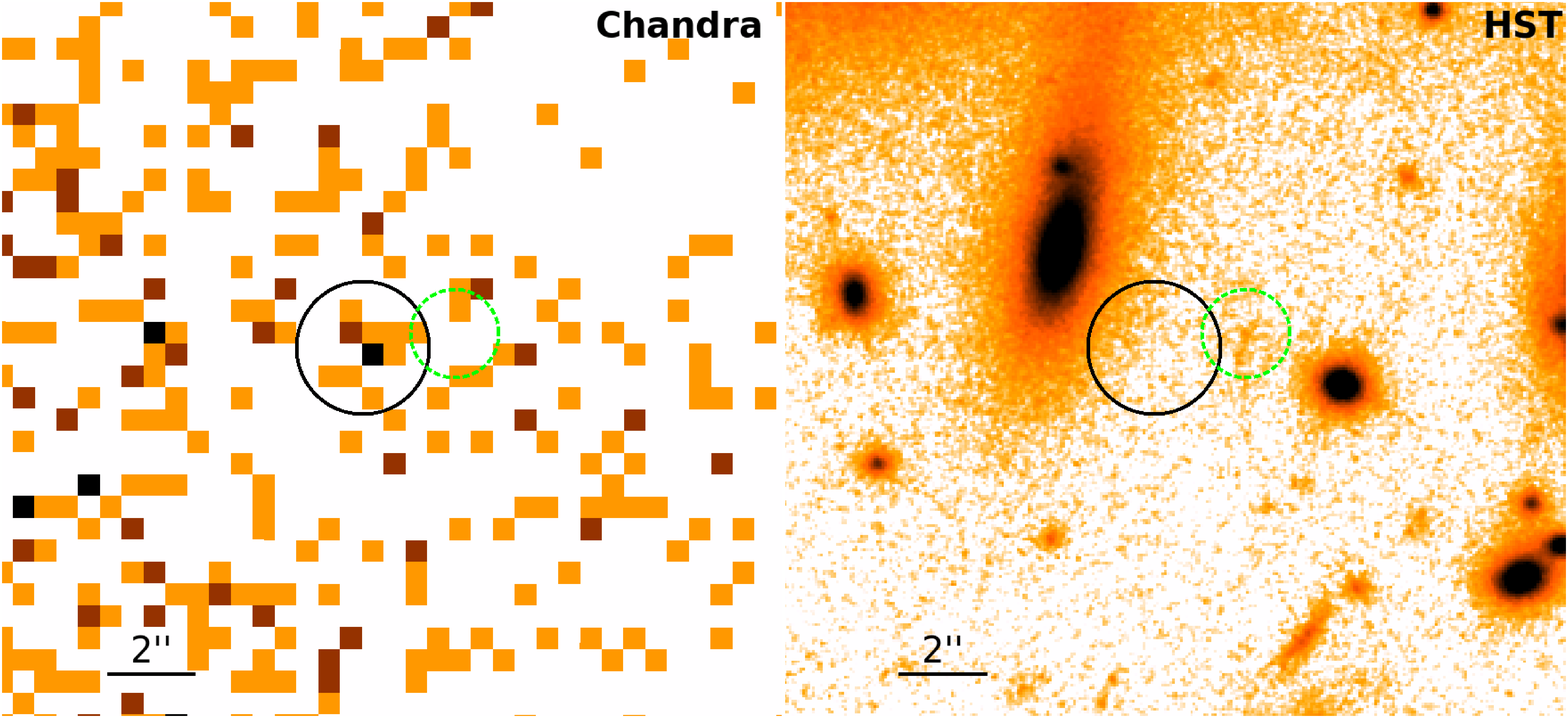} 
      \epsfxsize=0.85\textwidth \epsfbox{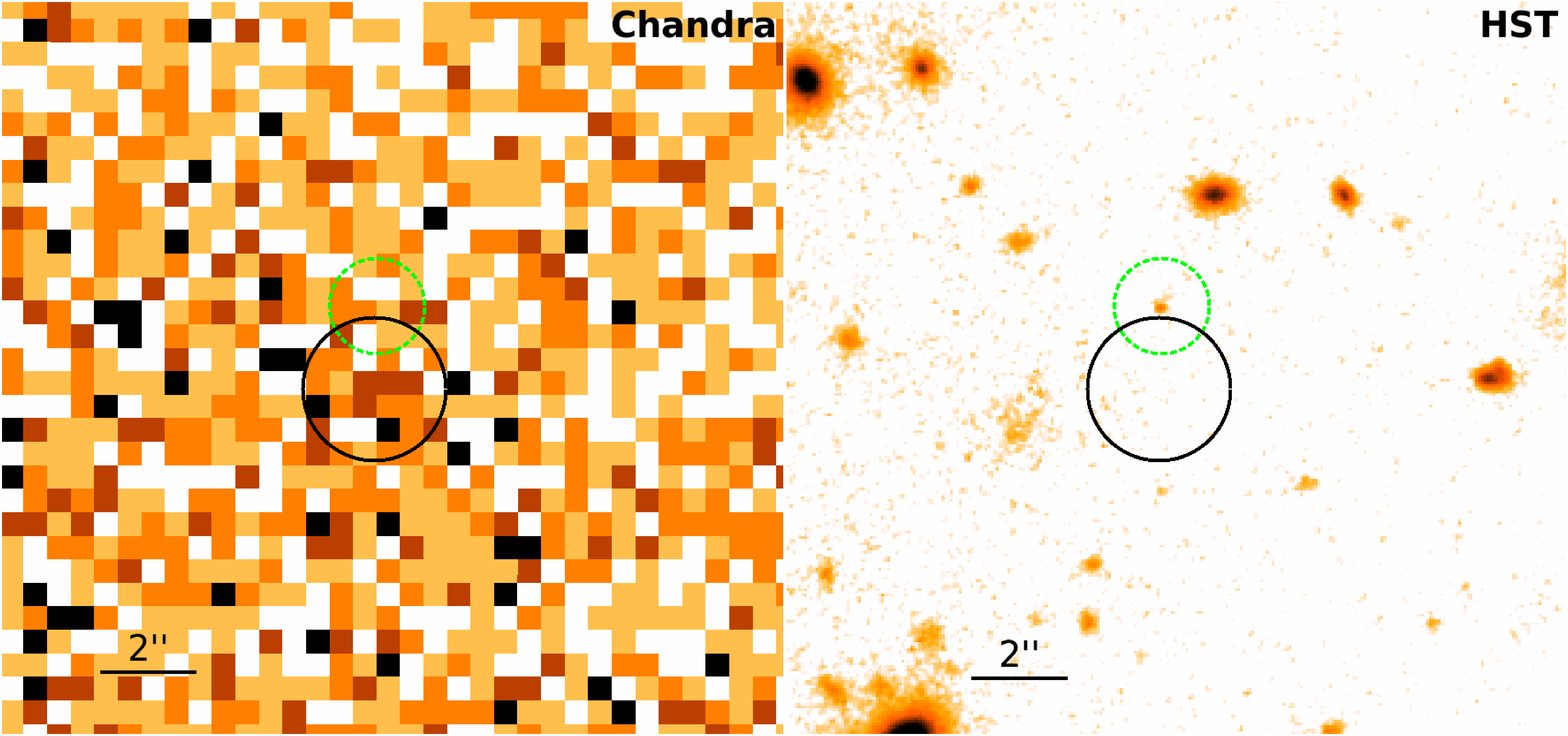}
      \vspace{0cm}
      \caption{The potential matches between the X-ray sources and high-redshift AGN for MACS0553-33 (top panel) and PLCKG287+32 (bottom panel). The left panels show the  $0.5-7$~keV band  \textit{Chandra} images and the right panels show the multi-color HST images of the regions around the sources. The images are centered on the X-ray sources (black solid circles) that are in the vicinity of galaxies at $z=6.55$ and $z=7.82$ (dashed green circles). The projected distances between the centroids of the X-ray point sources and the high-redshift galaxies are $2.1\arcsec$ and $1.7\arcsec$ for the source in MACS0553-33 and PLCKG287+32, respectively. However, due to the relatively large projected distance, the X-ray sources are unlikely to be associated with the high-redshift galaxies.} 
     \label{fig:matches}
  \end{center}
\end{figure*}

Although our main goal is to identify high-redshift AGN, it is expected that there will be many lower redshift AGN in the field-of-view that could contaminate the sample. To identify point sources in the individual galaxy clusters, we used the CIAO \textsc{wavdetect} tool. We searched for point sources in the merged images in the $0.5-7$~keV (broad band), $0.5-2$~keV (soft band), and $2-7$~keV (hard band) energy ranges. We derived point spread function (PSF) maps for each observation using the \textsc{mkspfmap} tool, which is used by  \textsc{wavdetect} to look up the size of the PSF for each pixel in the images. To detect both small and more extended features in the images, we used the square root two series of 2 from $\sqrt2$ to 16. In addition, we set the \textsc{ellsigma} parameter to 4, which assures that $\gtrsim90\%$ fraction of counts associated with the point sources are encircled. Therefore, the residual counts from point sources are not expected to significantly contribute to the large-scale diffuse emission. The significance threshold of the source detection was set to $10^{-6}$, which is expected to result in one false detection per each $1024\times1024$ pixel image. We identified 3558 point sources within the footprint of the galaxy clusters. Given the applied significance threshold of the \textsc{wavdetect} tool and the area covered by the detectors, we estimate that about $4\%$ of the sources are of spurious nature. This source list was used to probe whether individual high-redshift AGN at the known coordinates of lensed galaxies are detected in the \textit{Chandra} images. Finally, we excluded the detected point sources when carrying out the stacking analysis. 

We extracted images in the broad, soft, and hard bands. These images were used to  cross-correlate the X-ray source list with the galaxy positions and to stack the individual galaxies. To carry out the stacking, we cut out $100 \times 100$ pixel regions of the images and exposure maps around each galaxy, and co-added the image and exposure map cutouts. 

The lensing magnification affects the fluxes that we observe from the high-redshift AGN. Since the gravitational lensing is achromatic, photons in the optical and X-ray waveband are magnified by the same factor. To account for the magnification factors, we rely on the average lensing magnifications derived by  \citet{strait20}, who used a substantial set of lensing models to derive the median magnification factor at the individual galaxy's position and redshift (see Section \ref{sec:relics}). Because the lensing magnification only affects the emission from the source and not from the instrumental or sky background, we applied the magnification correction on the background subtracted count rates. Specifically, we multiplied each exposure map associated with the individual galaxies with the corresponding median lensing magnification taken from \citet{strait20}. To compute the fluxes and upper limits of the stacked AGN, we used the exposure maps, which were convolved with the lensing magnification.

\begin{table*}[!t]
\begin{center}
\caption{Individually detected X-ray sources with high-redshift galaxies in their proximity}
\begin{minipage}{18cm}
\renewcommand{\arraystretch}{1.3}
\centering
\begin{tabular}{ cccccccccc}
\hline
Cluster name & Galaxy ID& RA\textsubscript{X} & Dec\textsubscript{X} & RA\textsubscript{gal}  &  Dec\textsubscript{gal}  & $z_{\rm gal}$ & Offset & $F_{\rm 0.5-7keV}$ & $L_{\rm 0.5-7keV}$ \\
& & (J2000) & (J2000)&(J2000) & (J2000)& &  ($\arcsec$) & ($\rm{erg \ s^{-1} \ cm^{-2}}$)  & ($\rm{erg \ s^{-1} }$) \\
(1) & (2) & (3) & (4) & (5) & (6) & (7) & (8)& (9) & (10) \\
\hline
MACS0553-33 & 830 &05:53:18.40  &   $-$33:42:38.95    &    05:53:18.23     &    $-$33:42:38.63    &  6.55 &  $2.1$  &   $ 9.0\times10^{-17}$  & $ 4.6\times10^{43}$ \\  
PLCKG287+32 & 792 &   11:50:48.22  & $-$28:04:19.62     & 11:50:48:22     &  $-$28:04:17.89    &  7.82  &  $1.7$  &    $ 4.6\times10^{-16}$    &    $ 3.5\times10^{44}$     \\  
 \hline
\end{tabular} 
\end{minipage}
\end{center}
Columns are as follows. (1) Name of the RELICS cluster; (2) ID number in the RELICS high-redshift galaxy catalog (https://victoriastrait.github.io/relics); (3) and (4) Right ascension and declination of the X-ray source; (5) and (6) Right ascension and declination of the high-redshift galaxy in the proximity of the X-ray source; (7) Redshift of the galaxy; (8) Offset between the X-ray source and the high-redshift galaxy; (9) and (10) Flux and X-ray luminosity of the X-ray sources. To compute the luminosity, we assumed that the X-ray source is located at the redshift of the galaxy given in column (7). 
\label{tab:sources}
\end{table*}

\section{Results}
\label{sec:results}

\subsection{Individual detections}
\label{sec:detections}

We first investigated whether the merged \textit{Chandra} images of individual RELICS clusters can detect an AGN associated with the high-redshift galaxies. To this end, we cross-correlated the coordinates of the detected X-ray sources with those of the lensed galaxies. To maximize the likelihood of the search, we carried out the cross-correlation in all three energy ranges. We searched for counterparts within $2.5\arcsec$ radius. This fairly large search radius was chosen to conservatively account for any differences between the astrometric accuracy of \textit{HST} and \textit{Chandra} \citep{liu21} and to consider the broader \textit{Chandra} point spread function at the edges of the detectors. 

We identified two X-ray sources, which are in the proximity of high-redshift lensed galaxies. The galaxies are located in MACS0553-33 and PLCKG287+32 and their IDs in the high-redshift galaxy catalog\footnote{The properties of high-redshift galaxies in \citet{strait20} are listed at https://victoriastrait.github.io/relics} are 830 and 792, respectively. The redshifts of these galaxies are $z=6.55$ and $z=7.82$ for IDs 830 and 792. The coordinates and properties of the X-ray source-lensed galaxy pairs are given in Table \ref{tab:sources}. The offsets between the positions of the X-ray sources and the lensed galaxies are $2.1\arcsec$ and $1.7\arcsec$ for MACS0553-33 and PLCKG287+32, which are close to the upper limit of our search radius. Interestingly, the median lensing magnifications at the position of the galaxies are among the largest in our sample: it is $\mu = 16.87$ for  the source in MACS0553-33 and $\mu = 9.83$ for PLCKG287+32, respectively. 

In Figure \ref{fig:matches}, we show the broad band \textit{Chandra} and \textit{HST} images of the two X-ray sources and the high-redshift galaxies. The sources in MACS0553-33 and PLCKG287+32 have $6.5 \pm 3.6$ and $16.0 \pm 7.2$ net counts in the broad band, respectively. Both sources lie relatively close to the aim point, implying a narrow PSF. The offset between the X-ray source-galaxy pairs is about $2\arcsec$. At the redshift of the galaxies, the projected offset corresponds to $\sim10$~kpc, which is nearly an order of magnitude larger than the typical half light radius of typical galaxies at $z\sim6$ \citep{bouwens04}. Astrometric uncertainties cannot explain the large offset since the typical offset between  \textit{Chandra} and \textit{HST} astrometry is $ 0.35\arcsec $ \citep{liu21}. This suggests that the X-ray sources are not associated with the high-redshift galaxies.

Given the large number of X-ray sources detected on the  \textit{Chandra} images, it is possible that the association is due to coincidence. To estimate the likelihood of random matches, we carried out Monte Carlo simulations. We generated random coordinates within the footprint of the galaxy clusters. For each cluster, the number of random coordinates was the same as the number of high-redshift galaxies in the given cluster. Based on these sets of coordinates, we searched for matches between the simulated ``high-redshift galaxies'' and the X-ray sources. To obtain a statistically meaningful sample, we generated 10,000 random sets of ``high-redshift'' galaxies, which amounts to $155 \times 10^4 = 1.55 \times 10^6$ randomly selected coordinates. We identified 2,373 random matches for all these coordinates, which suggests that the likelihood of chance coincidence is $\approx0.15\%$. Thus, we expect $\approx0.23$ random matches. While this number is lower than two X-ray source-galaxy pairs, it is also possible that the X-ray sources are spurious detections (Section \ref{sec:chandra}).

The other X-ray sources in our sample do not have a high-redshift galaxy in their proximity. This suggests that most point sources are associated with AGN at lower redshifts or they may be foreground objects.

\begin{figure}[!]
  \begin{center}
    \leavevmode
    \hspace{0.0cm}
      \epsfxsize=0.48\textwidth \epsfbox{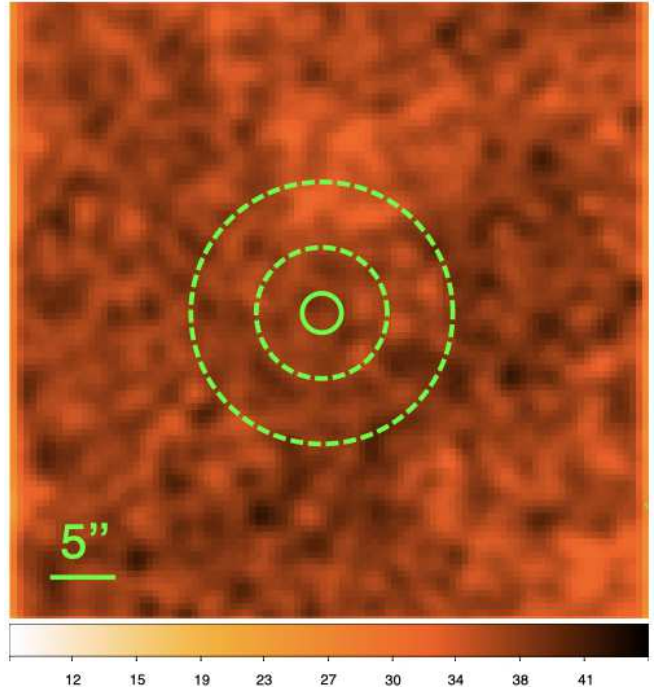}
      \vspace{0cm}
      \caption{Stacked $0.5-7$ keV band image of the 155 galaxies in our sample. The stacked sample does not reveal a statistically significant detection. The source and background regions used to derive the upper limits are shown with the circle (solid line) and annulus (dashed lines).} 
     \label{fig:allstack}
  \end{center}
\end{figure}

\begin{figure*}[!]
  \begin{center}
    \leavevmode
      \epsfxsize=0.75\textwidth \epsfbox{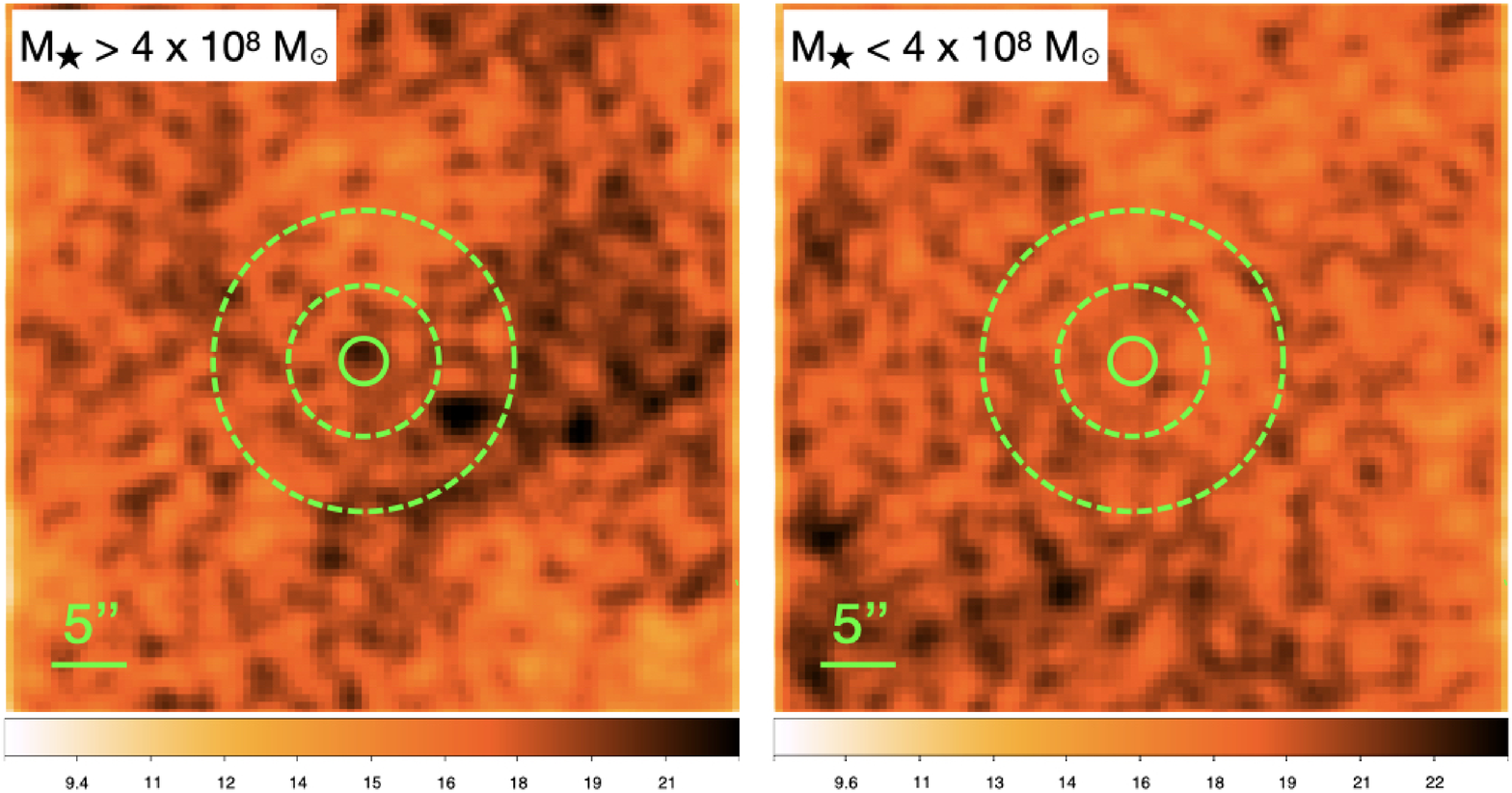}
      \epsfxsize=0.75\textwidth \epsfbox{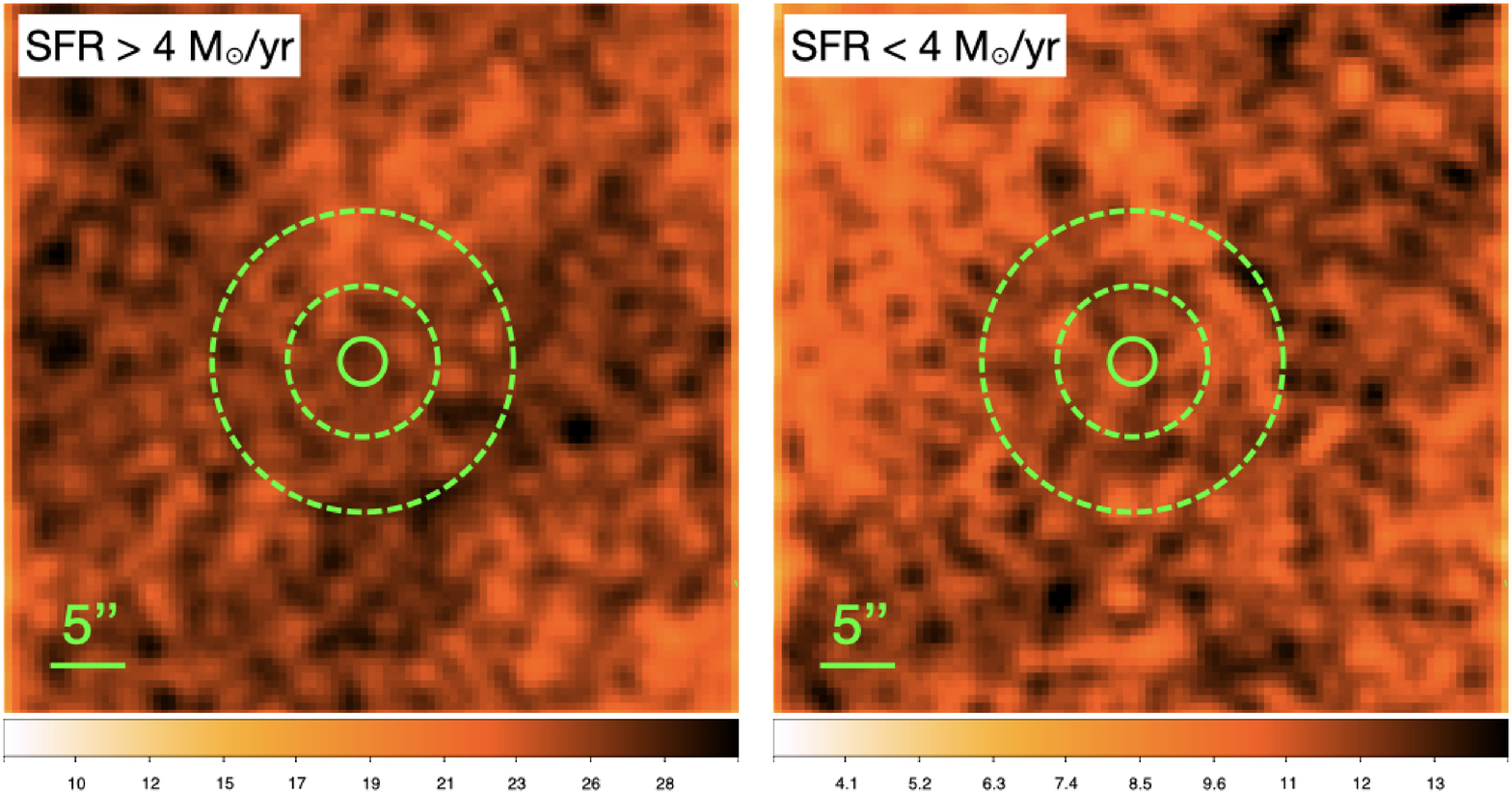}
      \epsfxsize=0.75\textwidth \epsfbox{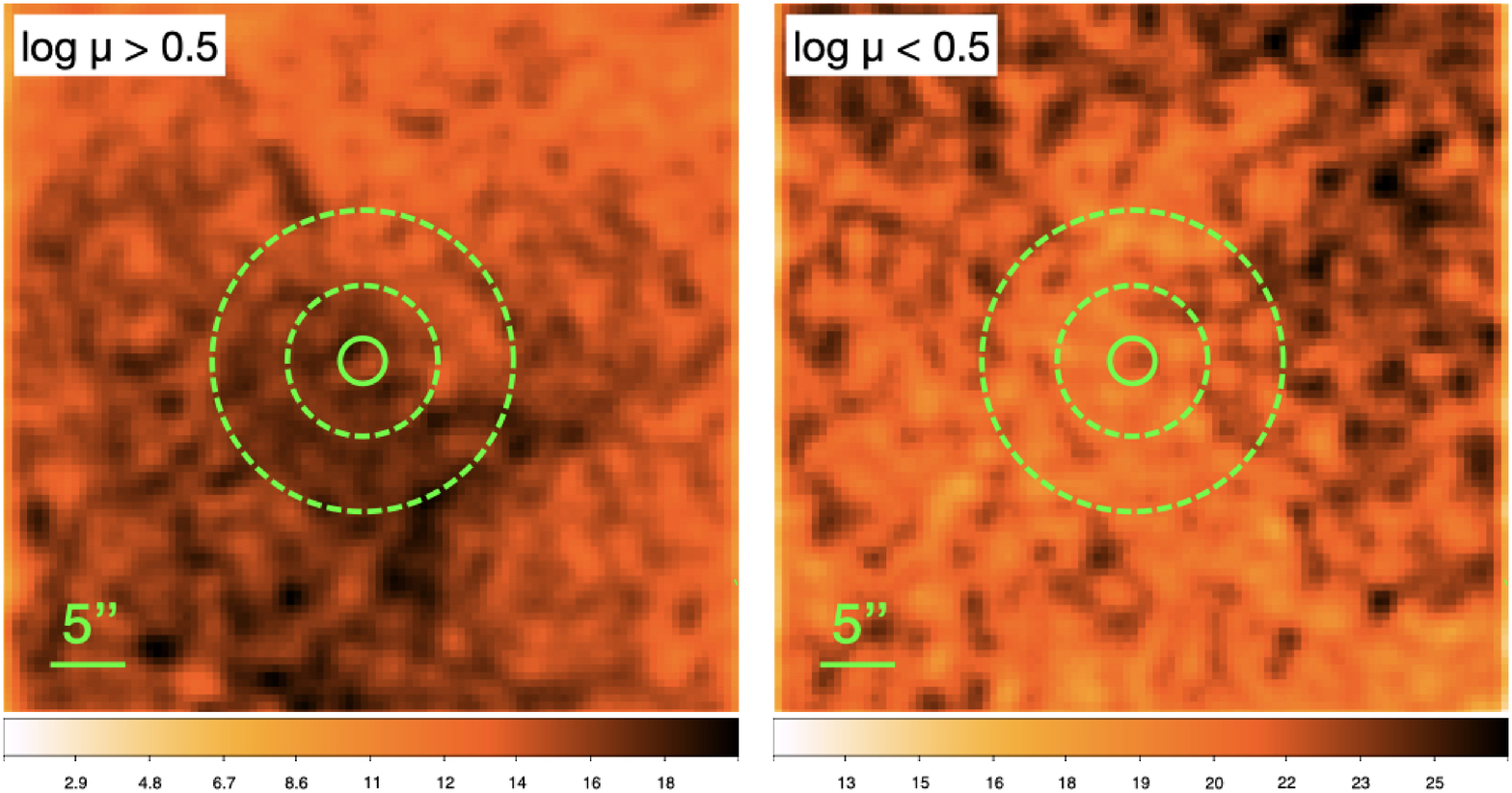}
      \vspace{0cm}
      \caption{Stacked $0.5-7$~keV band \textit{Chandra} images of lensed high-redshift galaxies using different binning criteria. We split the galaxies based on stellar mass (top row), star-formation rate (middle), and lensing magnification (bottom). We obtained a weak, $2.2\sigma$ detection for the high-mass sub-sample, while other sub-samples remained undetected. The physical criteria to divide the galaxies into bins is shown on the stacked images.} 
     \label{fig:stacks}
  \end{center}
\end{figure*}

\subsection{Stacking the high-redshift galaxies}
\label{sec:stacking}

To increase the signal-to-noise ratios and the likelihood of detecting high-redshift AGN, we stacked the X-ray photons associated with the individual galaxies. This approach, combined with the lensing magnifications, allows us to probe relatively faint AGN. Indeed, the sensitivity of our study is nearly compatible with that achieved in the 7~Ms Chandra Deep Field South regions (see Section \ref{sec:cdfs}). 

A major difference between the present work and previous stacking analyses  \citep[e.g.][]{vito16} is that we are co-adding galaxies that reside behind rich galaxy clusters. Therefore, emission from the intracluster medium (ICM) contributes to the overall emission and elevates the background level. In addition, the ICM exhibits notable structure across galaxy clusters, particularly in merging systems. In our analysis, we do not specifically account for the varying level of ICM emission associated with the individual lensed galaxies. Because the regions that were cut around individual galaxies are small (Section \ref{sec:chandra}) relative to the angular size of the galaxy clusters, variations in the ICM emission are negligible on these small scales and average out across the stacked images. 

We carried out the stacking analysis in all three energy ranges using multiple approaches. First, we co-added all 155 galaxies in our sample. In Figure \ref{fig:allstack}, we present the broad band stacked X-ray images of the galaxies. Visual inspection of the galaxies does not reveal a bright point source at the center of the images. When investigating the images extracted in the soft and hard band, we also did not identify a point source. To constrain the flux associated with the stacked AGN, we extracted the counts from the stacked images and applied the stacked lensing-corrected exposure maps. The source and background regions were described with a circular region with  $1.5\arcsec$ radius and with an annulus with $5\arcsec-10\arcsec$ radii, respectively. We did not detect a statistically significant signal associated with the stacked high-redshift AGN. In the absence of a detection, we place an upper limit on the flux. To be consistent with previous works \citep[e.g.][]{vito16}, we report the upper limits in the $0.5-2$~keV energy range. Taking into account the lensing-corrected exposure maps, the upper limit on the flux is $F_{\rm 0.5-2keV} < 2.1\times10^{-18}  \ \rm{erg \ s^{-1} \ cm^{-2}} $, which corresponds to a luminosity of  $L_{\rm 0.5-2keV} < 9.1 \times10^{41}  \ \rm{erg \ s^{-1}} $  using the mean redshift of $z=6.11$. 

Although the entire galaxy sample does not reveal a detection, we further split our sample to increase the likelihood of a detection. We used three criteria to divide our sample. First, we split the galaxies based on their stellar mass. It is well established that stellar bulge mass of galaxies is proportional with the BH mass in the local Universe \citep[e.g.][]{mcconnell13,saglia16}. Recently, this correlation was investigated for dwarf galaxies and it was found that the $M_{\rm BH} - M_{\rm bulge}$ relation can be extended to these low mass galaxies \citep{schutte19}. While this relation may be different for galaxies at $z\sim6$, it is reasonable to assume that more massive galaxies, even at the low-mass regime, host more massive BHs, which, in turn, are expected to shine brighter as AGN. Hence, by dividing our sample by mass, we expect that the average luminosity of AGN in the high-mass sample will become detectable. The luminosity of AGN in low-mass galaxies will be lower than that of the entire sample, implying that this sample will remain undetected. To split the sample based on mass, we applied $4\times10^8 \ \rm{M_{\rm \odot}}$ as our threshold, which approximately splits our sample into two equal-sized groups. After stacking the galaxies in the two sub-samples, we obtained a $2.2\sigma$ detection in the high-mass sample (Figure \ref{fig:allstack} top left panel), while the low-mass sample remained undetected (Figure \ref{fig:stacks}). Specifically, we detected $59.4\pm26.5$ net counts in the broad band, which corresponds to a weak detection. We note that the level of the detection significance does not exhibit notable variation if different background regions are used. We derive fluxes of $F_{\rm 0.5-2keV} = (8.1\pm3.6)\times10^{-18}  \ \rm{erg \ s^{-1} \ cm^{-2}} $ and $F_{\rm 0.5-2keV} < 2.5\times10^{-18}  \ \rm{erg \ s^{-1} \ cm^{-2}} $ for the two samples, which correspond to $L_{\rm 0.5-2keV} = (3.7\pm1.6) \times10^{42}  \ \rm{erg \ s^{-1}} $ and $L_{\rm 0.5-ke2V} < 1.0 \times10^{42}  \ \rm{erg \ s^{-1}} $. We further discuss this possible detection in Section \ref{sec:discussion}. 

The second approach to split the galaxies into two samples was based on the star formation rate. It is believed that the stellar population of galaxies co-evolves with their central BH as they are feeding from the same gas supply \citep[e.g.][]{hopkins06}. This, in turn, suggests that galaxies with high star formation rate may host BHs that are growing at a more rapid pace hence are more luminous. Additionally, galaxies with high star-formation rates are expected to host a more numerous population of high-mass X-ray binaries, whose total luminosity is proportional to the star formation rate of the host galaxy \citep{mineo12,fragos13}. Finally, actively star-forming galaxies are also more likely to host ultraluminous X-ray sources, which can have luminosities comparable to low-luminosity AGN \citep{kovlakas20}. Taken all together, galaxies with high star-formation rate may exhibit a higher X-ray luminosity and could be detected in the stacked sample. We split our galaxy sample into two groups using $4 \ \rm{M_{\odot} \ yr^{-1}}$ as the threshold. We did not detect a statistically significant signal from either sub-samples. Therefore, we derived $1\sigma$ upper limits on the fluxes and luminosities of the galaxies.  

As the third approach, we split the galaxy sample based on their median lensing magnification. The advantage of this method is that the lensing magnification only boosts the signal from the high-redshift galaxies (i.e.\ the luminosity of the AGN), while it does not increase the galaxy cluster emission and the overall background level. Hence, the high lensing magnification sample should result in improved signal-to-noise ratios from the AGN. We selected $\log \mu=0.5$ as our threshold to split the galaxies. After stacking the two sub-samples and measuring the count rates, we found that none of the two sub-samples exhibit detections. Therefore, we derived upper limits on the flux and luminosity of the AGN. 

We present the stacked $0.5-7$~keV band images in Figure \ref{fig:stacks} and we tabulated the results of the stacking analysis in Table \ref{tab:limits}. This table includes the number of galaxies, fluxes, and luminosities for each sub-sample. While we relied on Method A to split the galaxies into sub-samples, for completeness we also list the mean physical properties of the galaxies obtained through Method B. The results are further discussed in Section \ref{sec:discussion}.

\begin{table*}[!t]
\caption{Properties of the stacked samples}
\begin{minipage}{1.0\textwidth}
\renewcommand{\arraystretch}{1.3}
\centering
\begin{tabular}{ccccccccccc}
\hline
Bin &  $N_{\rm gal}$ & $z $ & $ \mu$ & $M_{\rm \star}$ (\#A) &
$M_{\rm \star}$ (\#B)  & SFR (\#A) & SFR (\#B) &  $F_{\rm obs,0.5-2keV}$ & $L_{\rm obs,0.5-2keV}$   \\
            & & & & ($\rm{M_{\rm \odot}}$) & ($\rm{M_{\rm \odot}}$) &  
($\rm{M_{\rm \odot} \ yr^{-1}}$) &  ($\rm{M_{\rm \odot} \ yr^{-1}}$) &
($\rm{erg \ s^{-1} \ cm^{-2}}$) & ($\rm{erg \ s^{-1} }$)  \\
(1) & (2) & (3) &(4) & (5) & (6) &(7) & (8) & (9) & (10) \\
\hline
High-mass & 73 & 6.22 & 3.6 & $ 2.6\times 10^9$ & $ 4.7\times 10^9$ & 16.8 & 22.0 & $(8.1\pm3.6) \times 10^{-18}$ & $(3.7\pm1.6)\times10^{42}$ \\
Low-mass & 82 & 6.01 & 7.9 & $ 1.4\times 10^8$ & $ 9.2\times 10^8$ & 4.4 & 3.8 &  $<2.5 \times 10^{-18}$ & $<1.0\times10^{42}$  \\
High-SFR & 105 & 6.17 & 5.8 & $ 1.8\times 10^9$ & $ 3.6\times 10^9$ & 13.7 & 16.7 &  $<2.7 \times 10^{-18}$ & $<1.1\times10^{42}$ \\
Low-SFR& 50 & 5.99 & 6.2 & $ 2.6\times 10^8$ & $ 8.3\times 10^8$ & 3.0 & 3.2 &  $<3.3 \times 10^{-18}$ & $<1.4\times10^{42}$ \\
$\log \mu >0.5$ & 48 & 6.37 & 13.8 & $ 1.3\times 10^9$ & $ 3.1\times 10^9$ & 10.6 & 17.3 &  $<2.6 \times 10^{-18}$ & $<1.2\times10^{42}$ \\
$\log \mu <0.5$ & 107 & 5.99 & 2.4 & $ 1.3\times 10^9$ & $ 2.5\times 10^9$ & 10.1 & 10.1 & $<3.4 \times 10^{-18}$ & $<1.4\times10^{42}$ \\
All RELICS & 155 & 6.11 & 5.91 & $ 1.3\times 10^9$ & $ 2.7\times 10^9$ & 10.3 & 12.3 &  $<2.1 \times 10^{-18}$ & $<9.1\times10^{41}$\\
\hline 
All CDF-S $^\dagger$ & 230 & 5.93  & -- & \multicolumn{2}{c}{$(0.2-60.1)\times10^9$} & \multicolumn{2}{c}{$(1.1-328.7)$}  &  $<4.4 \times 10^{-19}$ & $<1.8\times10^{41}$ \\
\hline
\end{tabular}
\vspace{0.1in}
\end{minipage}
$^\dagger$ All galaxies in the CDF-S field with $5.5<z<6.5$. The stellar mass and star formation rates for the CDF-S galaxies are based on \citet{santini15}. Because these values were \textit{not} computed following the method described in \citet{strait20}, we opt to show the range of these parameters. \\ Columns are as follows. (1) Binning method; (2) Number of galaxies in the bin; (3) Mean redshift of galaxies; (4) Median lensing magnification factor; (5) and (6) Mean stellar mass of galaxies obtained via Methods A and B in \citet{strait20}, respectively; (7) and (8) Mean star-formation rate of galaxies computed via Methods A and B in \citet{strait20}; (9) and (10) Observed flux and luminosity of the galaxies in the $0.5-2$~keV band, respectively.
\label{tab:limits}
\end{table*}

\section{Discussion}
\label{sec:discussion}

\subsection{Comparison with CDF-S}
\label{sec:cdfs}

So far the most powerful constraints on the average properties of high-redshift AGN were obtained by \citet{vito16}, who carried out a stacking analysis using the deep, 7~Ms, observations of the Chandra Deep Field South field. \citet{vito16} stacked AGN in three redshift bins: $3.5<z < 4.5$, $4.5<z< 5.5$, and $5.5<z<6.5$. They obtained a statistically significant detection in the lowest redshift bin, a tentative detection in the $4.5<z< 5.5$ bin, but AGN in the highest redshift bin remained undetected. Due to the large galaxy sample and the deep observations, the total stacked exposure time of galaxies in the $5.5<z < 6.5$ redshift bin was about $1.35\times10^9$~s and the derived flux upper limit was $<4.4\times10^{-19} \ \rm{erg\ s^{-1} \ cm^{-2}}$, which corresponds to a luminosity of $<1.8\times10^{41} \ \rm{erg \ s^{-1}}$ assuming $z=6$ (Table \ref{tab:limits}). 

Although the number of high redshift galaxies in our sample is similar to that studied in \citet{vito16}, the stacked exposure time of the galaxies in the RELICS sample is nearly two orders of magnitude shorter and is about $2.0 \times 10^7$~s. However, the lensing magnification of the galaxy clusters significantly improves our signal-to-noise ratios. Since the gravitational lensing is achromatic, it increases the signal from the high-redshift galaxies but not the background emission (including the X-ray emission originating from the intracluster medium). 

Similarly to the galaxies in CDF-S, we also did not detect a statistically significant signal from high-redshift galaxies when stacking all galaxies in the sample. The $0.5-2$~keV band flux upper limits are about a factor of $\sim4$ times higher than those obtained in the CDF-S footprint. Similarly to \citet{vito16}, we divided our sample based on the stellar mass of the host galaxies. Interestingly, we obtained a weak ($\sim2.2\sigma$) detection in the stacked galaxy sample, with fluxes of $8.1\times10^{-18} \ \rm{erg \ s \ cm^{-2}}$, which exceeds the CDF-S flux upper limit by factor of $\sim8$. However, due to the relatively low signal-to-noise ratio of the detection, we cannot exclude that this detection is an upward fluctuation. This detection is further discussed in the next Section.

\subsection{Detection in the high-mass sub-sample}
\label{sec:detection}

Our sample of massive high-redshift galaxies exhibited a $2.2\sigma$ detection. The corresponding  X-ray flux is factor of about 8 times higher than the upper limit obtained for $z\sim6$ galaxies in the CDF-S field. This result may appear controversial, especially when comparing the stellar masses of the galaxies in the CDF-S and RELICS samples. The median stellar mass of the RELICS galaxies is lower than that of the CDF-S galaxies (Table \ref{tab:limits}). However, this difference is likely due to systematic effects associated with the determination of stellar masses at high redshifts.

The SED fitting of the CDF-S galaxies was done by ten independent teams \citep{santini15}. The stellar masses presented in \citet{vito16} correspond to the median of the stellar masses obtained from these ten different groups. At $z<5.5$, the stellar mass estimates obtained through different methods are in good agreement with each other. However, at high redshift ($5.5<z<6.5$) there are large differences, which are mostly caused by the inclusion (or omission) of nebular line emission in the models. Specifically, only three teams included the nebular line emission in their SED fitting procedure, which, however, is essential to accurately fit the SED of high redshift galaxies. Indeed, emission lines, such as H-$\rm{\beta}$ and [OIII] can substantially contaminate the observed rest-frame window and drastically alter the SED fitting. The net effect of this is that the stellar mass of high redshift galaxies can be over-estimated. According to \citet{santini15}, the omission of nebular lines in young and/or high-redshift galaxies can result in stellar masses that are overestimated by up to a factor of 25. Based on \citet{santini15}, we estimate that not accounting for nebular emission lines will result in factor of $\sim3$ too high stellar masses for the overall $5.5<z<6.5$ galaxy population with $\sim15\%$ of the galaxies having a factor of $\sim5$ too high stellar masses. Since the stellar masses presented in \citet{santini15} are the median\footnote{The median of the stellar masses in \citet{santini15} was computed using the HodgesÐLehmann estimator.} of the values obtained by the different teams and most of the teams did not include emission from the nebular emission lines, the stellar masses for $5.5<z<6.5$ galaxies presented in \citet{santini15} and employed in \citet{vito16} are over-estimated.

\begin{figure}[!t]
  \begin{center}
    \leavevmode
    \hspace{-0.2cm}
      \epsfxsize=0.48\textwidth \epsfbox{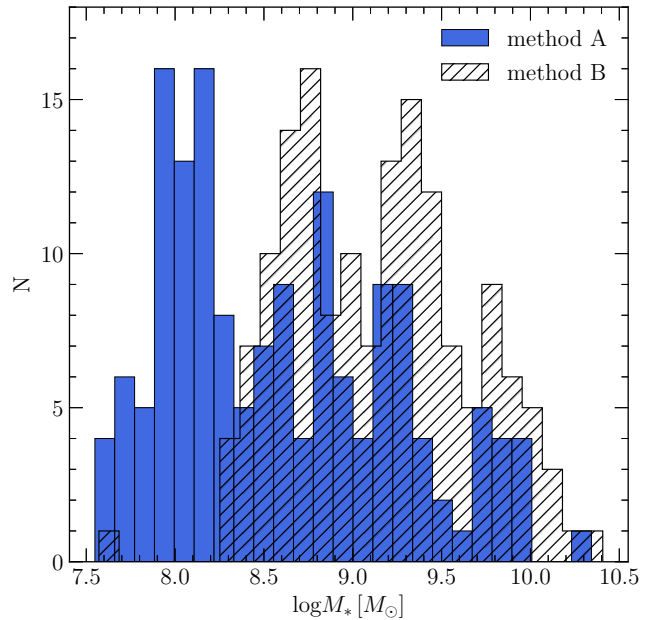}
      \vspace{0cm}
      \caption{Stellar mass distribution of the 155 high-redshift galaxies obtained using the two different SED fitting methods described in \citet{strait20}. In this work, we rely on the stellar masses derived using their Method A. The stellar masses obtained through Method B are systematically higher. Note that both SED fitting methods include nebular line emission.} 
     \label{fig:stmasses}
  \end{center}
\end{figure}

As opposed to this, nebular emission lines were included in the SED fitting procedure for the RELICS galaxies \citep{strait20}, which therefore provide more accurate stellar masses. Thus, the large offset between the stellar masses of the CDF-S and RELICS samples can be explained with the different SED fitting methods, in particular by the inclusion of nebular emission lines. Taking this difference into account, it is feasible that the RELICS high-mass sample includes -- on average -- more massive galaxies than the CDF-S field. In addition, we note that in the present work we utilize the stellar masses obtained through Method A in \citet{strait20}. However, the stellar masses derived through Method B are systematically higher than those in Method A; for example for the high-mass sample the stellar masses are factor of $\sim2$ higher (Figure \ref{fig:stmasses}). Therefore, the stacked detection of high-redshift galaxies behind the RELICS clusters is not incompatible with the non-detection of high-redshift galaxies in the CDF-S field. Indeed, the analysis of \citet{vito16} pointed out for the $3.5<z<5.5$ stacked samples that the detected X-ray signal from AGN is most sensitive to the stellar mass of the host galaxies and the most massive galaxies dominate the signal.

To further probe whether the $2.2\sigma$ detection is caused by the more massive  nature of galaxies in the high-mass bin or other, randomly selected, galaxies can reproduce the observed signal, we carried out Jackknife resampling. To this end, we randomly selected the coordinates of 73 high-redshift galaxies (which is identical with the number of galaxies in the high-mass bin) from the sample, co-added the X-ray photons associated with them, and derived the detection significance in the $0.5-7$~keV energy range. We repeated this analysis $10^5$ times to obtain a statistically meaningful sample. The detection significances of the randomly selected and stacked galaxies shows a  normal distribution that has a peak detection significance distribution at $\approx0.16\sigma$. We found that only $\approx0.3\%$ of the random resampling simulations show $\geq2.2\sigma$ detections. This is less frequent  than that suggested by the $2.2\sigma$ detections, hinting that the  detection in the massive high-redshift galaxies is unlikely to be the result of chance coincidence.

\subsection{Comparison with individual detections}
\label{sec:individual}

Over the past decades, wide-area optical and infrared surveys identified a substantial population ($>200$) of extremely luminous quasars at $z>5.5$  \citep[e.g.][]{fan06,willott07,jiang08,mortlock11,venemans13,banados14,wang17,wang19,yang17,yang19,yang20}. A fraction of these sources were followed up using X-ray observatories, which allowed the determination of the X-ray properties of these high-redshift AGN \citep[e.g.][]{nanni17,vito19,pons20,wang21}. These studies established that the X-ray luminosity of these sources ranges from several times $10^{43} \ \rm{erg \ s^{-1}}$ to a few times $10^{45} \ \rm{erg \ s^{-1}}$. 

Clearly, the luminosity of the quasars studied in \citet{nanni17} surpasses the X-ray upper limits obtained in our stacking analysis by about $1.5-3.5$ orders of magnitude. Although the source detection sensitivity strongly varies in our galaxy cluster sample due to the different exposure times and spatially varying lensing magnifications, we estimate the typical detection sensitivity of our study. To this end, we rely on the average \textit{Chandra} exposure time ($t_{\rm exp} = 104$~ks) of the RELICS galaxy cluster sample (Table 1), use the median lensing magnification factor of $\mu = 3$, apply the median redshift ($z=5.9$) of the lense galaxies, and assume that an X-ray AGN can be detected with 10 counts. We thus obtain an average source detection sensitivity of $\sim10^{44} \ \rm{erg \ s^{-1}}$, which implies that the bulk of the quasars in the sample of \citet{nanni17} could be individually detected behind the RELICS clusters. 

In the sample of RELICS galaxies, we did not identify a conclusive X-ray source-galaxy pair. In the absence of detection, we place an upper limit on the number of luminous AGN in high redshift galaxies using the Bayesian formalism for Poisson-distributed data \citep{kraft91}. The $95\%$ confidence interval for non-detection yields an upper limit of 3 sources, which implies that  $<1.9\%$ of $z\gtrsim6$ galaxies host AGN with luminosities of $\gtrsim4\times10^{43} \ \rm{erg \ s^{-1}}$. This result is also in line with our stacking analysis, which suggests that no more than a few per cent of galaxies may host AGN with such high luminosities.

\subsection{High mass X-ray binaries}
\label{sec:hmxb}

Although the primary goal of our study is to constrain the X-ray emission from luminous high-redshift AGN, other sources also contribute to the overall X-ray emission. Most notably, high-mass X-ray binaries also present a major source of X-ray emission. The X-ray emission from these sources is proportional to the star formation rate of the galaxy. Taking into account the $\rm{L_{X}}$-SFR relation \citep{fragos13}, the redshift evolution of the relation \citep{lehmer16}, and assuming a power law slope with $\Gamma = 2$, the average expected $0.5-2$~keV band luminosity from HMXBs is $1.4 \times 10^{41} \ \rm{erg \ s^{-1}}$.  

This average HMXB luminosity is about an order of magnitude below our detection limit. Therefore, given the sensitivity of our stacking analysis, it is unlikely that HMXBs can be detected in the present data set, even in our high-SFR sub-sample. Therefore, the non-detection of X-ray emission from HMXBs is consistent with the estimated X-ray luminosity from these sources. As a caveat, we note that the $\rm{L_{X}}$-SFR scaling relation has only been probed up to $z\sim2.5$ and it cannot be excluded that $z\approx6$ galaxies exhibit a different relation. However, based on the observed redshift evolution of this relation, it seems unlikely that their contribution will be close to the sensitivity of our stacking analysis. Additionally, we note that the X-ray emission from HMXBs are also not expected to play a significant role in the obtained $2.2\sigma$ detection in the high-mass bin. Specifically, HMXB are expected to account for $\sim6\%$ of the total emission in the $0.5-2$~keV energy range.
  
The predicted X-ray luminosity associated with high-redshift galaxies with the highest star-formation rates ($\sim100\ \rm{M_{\odot} \ yr^{-1}}$) is about $1.5\times10^{42} \ \rm{erg \ s^{-1}}$. While these luminosities are comparable with the average sensitivity of the stacks, these values remain below the individual source detection limits, which is also consistent with the fact that none of the galaxies with high star-formation rate were individually detected on the \textit{Chandra} images. The metallicity of galaxies strongly influences the number of HMXBs \citep[e.g.][]{linden10,lehmer16}. Specifically, the number of HMXBs per unit stellar mass is significantly higher in galaxies with low ($\sim20\%$) metallicities \citep{douna15}, with the excess sources being mostly present at the bright end of the X-ray luminosity function \citep{lehmer21}. While the effect of decreasing metallicity at higher redshift galaxies was incorporated in the model of \citet{lehmer16}, galaxies with extremely low metallicities and high star-formation rates could be individually detected in other lensed high-redshift galaxies. Additionally, future, more sensitive X-ray surveys will be able to detect the X-ray emission from HMXBs in high-redshift galaxies.

\subsection{Constraining the properties of $z\sim6$ BHs}
\label{sec:bhmass}

Based on the data from the stacked galaxies, we can constrain the characteristics of the \textit{typical} $z\sim6$ AGN. The X-ray upper limit on the luminosity of the BHs in the RELICS galaxy sample is $L_{\rm{X}}<8.4\times10^{41} \ \rm{erg \ s^{-1}}$ (Table \ref{tab:limits}). Assuming a bolometric correction of $K_{\rm bol}=10$ \citep{lusso12,duras20}, the upper limit on the bolometric luminosity is $L_{\rm{bol}}<8.4\times10^{42} \ \rm{erg \ s^{-1}}$. If BHs accrete at their Eddington rate, this upper limit corresponds to a mean BH mass of $M_{\rm BH} < 6.7\times10^4 \ \rm{M_{\odot}}$. Clearly, this average BH mass is about 4 orders of magnitude lower than those detected in the most luminous quasars. 

If we assume that galaxies at $z\sim6$ obey the BH mass--bulge mass scaling relation obtained in the local Universe \citep{schutte19} and consider the mean stellar mass of our sample ($1.3\times10^9 \ \rm{M_{\odot}}$), we expect a mean BH mass of $M_{\rm BH} = 2.6\times10^6 \ \rm{M_{\odot}}$. This implies that the BH mass calculated from the mean stellar mass is $\sim40$ times larger than the upper limit assuming Eddington accretion. Thus, our results imply that BHs are either much less massive at $z\sim6$ than expected from the local scaling relation or they accrete at a few per cent of their Eddington rate. We note that the former possibility is incompatible with some observational studies. Specifically, \citet{merloni10} suggested that BHs at high redshift may be over-massive relative to their host galaxies, and \citet{bogdan12} hinted that some BHs may grow faster than their host galaxies at high redshift. However, the relatively low mean accretion rate is feasible if most BHs originate from the heavy seed scenario. In this picture, BHs may experience episodic periods with high accretion rates, while most times they accrete at low Eddington rates. Obscuration of AGN by gas and dust may also play an important role in the non-detection, as it could hide the X-ray emission from these sources (see Section \ref{sec:obscuration} for details). Finally, it is also feasible that the BH occupation fraction of galaxies is not 100\%, implying that a fraction of low-mass systems do not host a BH \citep{lippai09,bellovary11}.  

Performing a similar calculation for the high-mass sub-sample that has a weak X-ray detection results in the same conclusion. The bolometric X-ray luminosity for these BHs is $L_{\rm{bol}}=3.7\times10^{42} \ \rm{erg \ s^{-1}}$, which corresponds to a mean BH mass of $M_{\rm BH} = 3.0 \times10^5 \ \rm{M_{\odot}}$ assuming Eddington accretion. The average stellar mass of our high-mass sample is $2.6\times10^9 \ \rm{M_{\odot}}$, which would imply that the mean BH mass is $6.5 \times10^6 \ \rm{M_{\odot}}$ if they follow the local scaling relation. The latter value is $\sim22$ times lower than that obtained assuming the Eddington accretion rate. Therefore, it is likely that even these more massive BHs do not accrete at the Eddington limit, but the average accretion rate remains at the $\sim5\%$ level. 

The low average accretion rate is not surprising when compared to theoretical studies. Most of the BH accretion, and hence BH growth, is expected to happen in the most luminous AGN and it is expected that BHs with low accretion rates will have a small contribution to the overall BH growth. For example, \citet{volonteri16} established, based on the Horizon-AGN simulations, that at high-redshifts ($z>4$), the bulk of the BH growth takes place in the most luminous ($L_{\rm bol} > 10^{44} \ \rm{erg \ s^{-1}}$) and most rapidly accreting ($f>0.1f_{\rm Edd}$) AGN. Because the sensitivity of our stacking analysis is well below this limit, the non-detection of the AGN in the full stacked RELICS galaxy sample is consistent with this. Since the observed luminosity of AGN depends both on the Eddington ratio and the BH mass, the detection of AGN in the high-mass galaxy sub-sample may be attributed to the more massive BH sample even though the mean accretion rates are still low. 

\subsection{Obscuration}
\label{sec:obscuration}

A fraction of AGN remains hidden behind gas and dust that absorbs the emission from the accretion disk of the BH. Because the obscured fraction of AGN may depend on both the luminosity and the redshift, our understanding of the complete census and evolution of AGN depends on the obscured fraction of AGN. A wide range of studies explored the obscuration of  high-redshift AGN, which found that about $50\%$ of $3<z<6$ AGN are obscured by a column density $ N_{H} > 10^{23} \ \rm{cm^-2} $ \citep{vito13,vito18,hickox18}.

 To assess the importance of obscuration, we follow \citet{vito16} and assume that $50\%$ of the AGN are obscured by a column density of $ N_{H} = 3.2 \times 10^{23} \ \rm{cm^{-2}} $ and assume that all sources have the same intrinsic luminosity. Given these conditions, the transmission factor for typical $z=6$ AGN is $c=0.3$, yielding a correction factor of $F_{\rm 0.5-2keV,total}/F_{0.5-2keV,obs} = 1.3$. While this factor is not applied for the observed X-ray fluxes and luminosities, the total -- obscuration corrected -- fluxes and luminosities can be computed by using this correction factor.

\subsection{Outlook}
\label{sec:outlook}

This work represents the first attempt to utilize gravitational lensing to constrain the average characteristics of AGN in  $z\sim6$ galaxies. We obtained a  $2.2\sigma$ detection of galaxies in the high-mass sub-sample, while other sub-samples remained undetected. This initial result is encouraging: it emphasizes the powerful nature of our approach and highlights that future detections with higher statistical significance are feasible. 

To further advance our understanding of high-redshift AGN,  the advancement of both optical/infrared observatories and deeper X-ray observations are required. \textit{JWST} will completely revolutionize our understanding of the early Universe by detecting large samples of faint and high-redshift galaxies. Two instruments aboard \textit{JWST}, the Near InfraRed Camera (NIRCam)  and the Mid InfraRed Instrument (MIRI), will provide broadband photometry that will allow probing the rest-frame UV, optical, and near-infrared spectral energy distributions of high-redshift galaxies. Thanks to its large collecting area, superb angular resolution, and infrared sensitivity, \textit{JWST}  will explore a vast population of galaxies at $z=6-10$ and will even  identify the first galaxies at $z\sim15$. By utilizing existing, upcoming, and proposed deep \textit{Chandra} observations of lensed galaxy clusters, the sample of high-redshift AGN can be significantly increased, thereby improving the signal-to-noise ratios, which may lead to further detection of AGN and/or the population of HMXBs in distant galaxies, both individually and in stacks. 

On a longer timescale, more sensitive high-resolution X-ray telescopes could provide an edge in detecting high-redshift galaxies that were identified by \textit{JWST}. For example, the proposed \textit{Lynx} observatory would be able to reach sensitivities to even individually detect some of the lensed AGN and would be able to probe the average luminosity of HMXBs. Indeed,  \textit{Lynx}  could drastically change the landscape of high-redshift AGN studies as it will detect AGN with luminosities with $10^{41} \ \rm{erg \ s^{-1}}$  in deep fields and at even lower luminosities by utilizing the lensing magnification of galaxy clusters. While \textit{JWST} and \textit{Lynx} would not operate simultaneously, X-ray follow-up observations of \textit{JWST} targets will revolutionize our understanding about high-redshift AGN.

\section{Conclusions}
\label{sec:conclusions}

In this work, we analyzed \textit{Chandra} X-ray observations of 34 RELICS galaxy clusters and probed the X-ray emission originating from 155 gravitationally-lensed galaxies  that reside at $z\gtrsim6$. We probed the emission from high-redshift AGN both individually and in stacks.  Our results can be summarized as follows: 
\begin{itemize}
\item To search for individually detected AGN associated with the high-redshift galaxies, we cross-matched the coordinates of the detected X-ray sources with those of the lensed galaxies. We did not identify an X-ray source that was undoubtedly associated with a high-redshift galaxy. 
\item To probe the average X-ray luminosity from high-redshift galaxies, we co-added the X-ray signal from the 155 lensed galaxies, but did not obtain a statistically significant detection. In the absence of detection, we placed a flux upper limit on the high-redshift AGN, which was $<2.1\times10^{-18} \ \rm{erg \ s^{-1} \ cm^{-2}}$, which corresponds to a luminosity upper limit of $<8.4\times10^{41} \ \rm{erg \ s^{-1}}$. 
\item Assuming that $z\sim6$ galaxies follow the local BH mass--bulge mass scaling relation, we estimate that the typical BH mass is $\sim2.6\times10^6 \ \rm{M_{\odot}}$. Given the upper limit on the luminosity, this implies that typical BHs at high redshift accrete at $5-10\%$ of their Eddington rate. 
\item We split galaxies based on their stellar mass, star-formation rate, and lensing magnification and stacked galaxies in these sub-samples. We obtained a weak, $2.2\sigma$, detection for massive galaxies. Taken at face value, the luminosity of the AGN in the high-mass group is $3.7\times10^{42} \ \rm{erg \ s^{-1}}$.  We did not obtain a statistically significant detection in other sub-samples. 
\item We find that emission from HMXBs is well below the sensitivity of the stacking analysis given the star-formation rate of the sample. 
\end{itemize}

\bigskip

\begin{small}
\noindent
\textit{Acknowledgements.}
We thank the reviewer the constructive report that improved this manuscript. The scientific results reported in this article are based to a significant degree on data obtained from the Chandra Data Archive. This research has made use of software provided by the Chandra X-ray Center (CXC) in the application packages CIAO, ChIPS, and Sherpa. This work is based on observations taken by the RELICS Treasury Program (GO 14096) with the NASA/ESA HST, which is operated by the Association of Universities for Research in Astronomy, Inc., under NASA contract NAS5-26555. \'A.B., C.J., W.R.F, and R.K. acknowledge support from the Smithsonian Institution and the Chandra High Resolution Camera Project through NASA contract NAS8-03060. O.E.K is supported by the GA\v{C}R EXPRO grant No. 21-13491X. M.B. and V.S. acknowledge support  provided by NASA/HST grant HST-GO-15920. 
\end{small}

\bibliographystyle{apj}
\bibliography{paper2.bib} 

\begin{thebibliography}{}
\expandafter\ifx\csname natexlab\endcsname\relax\def\natexlab#1{#1}\fi

\bibitem[{{Ai} {et~al.}(2016){Ai}, {Dou}, {Fan}, {Wang}, {Wu}, \&
  {Bian}}]{ai16}
{Ai}, Y., {Dou}, L., {Fan}, X., {et~al.} 2016, \apjl, 823, L37

\bibitem[{{Ba{\~n}ados} {et~al.}(2014){Ba{\~n}ados}, {Venemans}, {Morganson},
  {Decarli}, {Walter}, {Chambers}, {Rix}, {Farina}, {Fan}, {Jiang}, {McGreer},
  {De Rosa}, {Simcoe}, {Wei{\ss}}, {Price}, {Morgan}, {Burgett}, {Greiner},
  {Kaiser}, {Kudritzki}, {Magnier}, {Metcalfe}, {Stubbs}, {Sweeney}, {Tonry},
  {Wainscoat}, \& {Waters}}]{banados14}
{Ba{\~n}ados}, E., {Venemans}, B.~P., {Morganson}, E., {et~al.} 2014, \aj, 148,
  14

\bibitem[{{Begelman} {et~al.}(2006){Begelman}, {Volonteri}, \&
  {Rees}}]{begelman06}
{Begelman}, M.~C., {Volonteri}, M., \& {Rees}, M.~J. 2006, \mnras, 370, 289

\bibitem[{{Bellovary} {et~al.}(2011){Bellovary}, {Volonteri}, {Governato},
  {Shen}, {Quinn}, \& {Wadsley}}]{bellovary11}
{Bellovary}, J., {Volonteri}, M., {Governato}, F., {et~al.} 2011, \apj, 742, 13

\bibitem[{{Bogd{\'a}n} {et~al.}(2012){Bogd{\'a}n}, {Forman}, {Zhuravleva},
  {Mihos}, {Kraft}, {Harding}, {Guo}, {Li}, {Churazov}, {Vikhlinin}, {Nulsen},
  {Schindler}, \& {Jones}}]{bogdan12}
{Bogd{\'a}n}, {\'A}., {Forman}, W.~R., {Zhuravleva}, I., {et~al.} 2012, \apj,
  753, 140

\bibitem[{{Bouwens} {et~al.}(2004){Bouwens}, {Illingworth}, {Blakeslee},
  {Broadhurst}, \& {Franx}}]{bouwens04}
{Bouwens}, R.~J., {Illingworth}, G.~D., {Blakeslee}, J.~P., {Broadhurst},
  T.~J., \& {Franx}, M. 2004, \apjl, 611, L1

\bibitem[{{Brammer} {et~al.}(2008){Brammer}, {van Dokkum}, \&
  {Coppi}}]{brammer08}
{Brammer}, G.~B., {van Dokkum}, P.~G., \& {Coppi}, P. 2008, \apj, 686, 1503

\bibitem[{{Carnall} {et~al.}(2018){Carnall}, {McLure}, {Dunlop}, \&
  {Dav{\'e}}}]{carnall18}
{Carnall}, A.~C., {McLure}, R.~J., {Dunlop}, J.~S., \& {Dav{\'e}}, R. 2018,
  \mnras, 480, 4379

\bibitem[{{Connor} {et~al.}(2020){Connor}, {Ba{\~n}ados}, {Mazzucchelli},
  {Stern}, {Decarli}, {Fan}, {Farina}, {Lusso}, {Neeleman}, \&
  {Walter}}]{connor20}
{Connor}, T., {Ba{\~n}ados}, E., {Mazzucchelli}, C., {et~al.} 2020, \apj, 900,
  189

\bibitem[{{Douna} {et~al.}(2015){Douna}, {Pellizza}, {Mirabel}, \&
  {Pedrosa}}]{douna15}
{Douna}, V.~M., {Pellizza}, L.~J., {Mirabel}, I.~F., \& {Pedrosa}, S.~E. 2015,
  \aap, 579, A44

\bibitem[{{Duras} {et~al.}(2020){Duras}, {Bongiorno}, {Ricci}, {Piconcelli},
  {Shankar}, {Lusso}, {Bianchi}, {Fiore}, {Maiolino}, {Marconi}, {Onori},
  {Sani}, {Schneider}, {Vignali}, \& {La Franca}}]{duras20}
{Duras}, F., {Bongiorno}, A., {Ricci}, F., {et~al.} 2020, \aap, 636, A73

\bibitem[{{Fan} {et~al.}(2006){Fan}, {Strauss}, {Richards}, {Hennawi},
  {Becker}, {White}, {Diamond-Stanic}, {Donley}, {Jiang}, {Kim}, {Vestergaard},
  {Young}, {Gunn}, {Lupton}, {Knapp}, {Schneider}, {Brandt}, {Bahcall},
  {Barentine}, {Brinkmann}, {Brewington}, {Fukugita}, {Harvanek}, {Kleinman},
  {Krzesinski}, {Long}, {Neilsen}, {Nitta}, {Snedden}, \& {Voges}}]{fan06}
{Fan}, X., {Strauss}, M.~A., {Richards}, G.~T., {et~al.} 2006, \aj, 131, 1203

\bibitem[{{Fragos} {et~al.}(2013){Fragos}, {Lehmer}, {Tremmel}, {Tzanavaris},
  {Basu-Zych}, {Belczynski}, {Hornschemeier}, {Jenkins}, {Kalogera}, {Ptak}, \&
  {Zezas}}]{fragos13}
{Fragos}, T., {Lehmer}, B., {Tremmel}, M., {et~al.} 2013, \apj, 764, 41

\bibitem[{{Gallerani} {et~al.}(2017){Gallerani}, {Zappacosta}, {Orofino},
  {Piconcelli}, {Vignali}, {Ferrara}, {Maiolino}, {Fiore}, {Gilli},
  {Pallottini}, {Neri}, \& {Feruglio}}]{gallerani17}
{Gallerani}, S., {Zappacosta}, L., {Orofino}, M.~C., {et~al.} 2017, \mnras,
  467, 3590

\bibitem[{{Giacconi} {et~al.}(2001){Giacconi}, {Rosati}, {Tozzi}, {Nonino},
  {Hasinger}, {Norman}, {Bergeron}, {Borgani}, {Gilli}, {Gilmozzi}, \&
  {Zheng}}]{giacconi01}
{Giacconi}, R., {Rosati}, P., {Tozzi}, P., {et~al.} 2001, \apj, 551, 624

\bibitem[{{Hickox} \& {Alexander}(2018)}]{hickox18}
{Hickox}, R.~C., \& {Alexander}, D.~M. 2018, \araa, 56, 625

\bibitem[{{Hopkins} {et~al.}(2006){Hopkins}, {Hernquist}, {Cox}, {Di Matteo},
  {Robertson}, \& {Springel}}]{hopkins06}
{Hopkins}, P.~F., {Hernquist}, L., {Cox}, T.~J., {et~al.} 2006, \apjs, 163, 1

\bibitem[{{Jiang} {et~al.}(2008){Jiang}, {Fan}, {Annis}, {Becker}, {White},
  {Chiu}, {Lin}, {Lupton}, {Richards}, {Strauss}, {Jester}, \&
  {Schneider}}]{jiang08}
{Jiang}, L., {Fan}, X., {Annis}, J., {et~al.} 2008, \aj, 135, 1057

\bibitem[{{Kovlakas} {et~al.}(2020){Kovlakas}, {Zezas}, {Andrews}, {Basu-Zych},
  {Fragos}, {Hornschemeier}, {Lehmer}, \& {Ptak}}]{kovlakas20}
{Kovlakas}, K., {Zezas}, A., {Andrews}, J.~J., {et~al.} 2020, \mnras, 498, 4790

\bibitem[{{Kraft} {et~al.}(1991){Kraft}, {Burrows}, \& {Nousek}}]{kraft91}
{Kraft}, R.~P., {Burrows}, D.~N., \& {Nousek}, J.~A. 1991, \apj, 374, 344

\bibitem[{{Lehmer} {et~al.}(2016){Lehmer}, {Basu-Zych}, {Mineo}, {Brandt},
  {Eufrasio}, {Fragos}, {Hornschemeier}, {Luo}, {Xue}, {Bauer}, {Gilfanov},
  {Ranalli}, {Schneider}, {Shemmer}, {Tozzi}, {Trump}, {Vignali}, {Wang},
  {Yukita}, \& {Zezas}}]{lehmer16}
{Lehmer}, B.~D., {Basu-Zych}, A.~R., {Mineo}, S., {et~al.} 2016, \apj, 825, 7

\bibitem[{{Lehmer} {et~al.}(2021){Lehmer}, {Eufrasio}, {Basu-Zych}, {Doore},
  {Fragos}, {Garofali}, {Kovlakas}, {Williams}, {Zezas}, \&
  {Santana-Silva}}]{lehmer21}
{Lehmer}, B.~D., {Eufrasio}, R.~T., {Basu-Zych}, A., {et~al.} 2021, \apj, 907,
  17

\bibitem[{{Linden} {et~al.}(2010){Linden}, {Kalogera}, {Sepinsky}, {Prestwich},
  {Zezas}, \& {Gallagher}}]{linden10}
{Linden}, T., {Kalogera}, V., {Sepinsky}, J.~F., {et~al.} 2010, \apj, 725, 1984

\bibitem[{{Lippai} {et~al.}(2009){Lippai}, {Frei}, \& {Haiman}}]{lippai09}
{Lippai}, Z., {Frei}, Z., \& {Haiman}, Z. 2009, \apj, 701, 360

\bibitem[{{Liu} {et~al.}(2021){Liu}, {Tozzi}, {Rosati}, {Bergamini}, {Bartosch
  Caminha}, {Gilli}, {Grillo}, {Meneghetti}, {Mercurio}, {Nonino}, \&
  {Vanzella}}]{liu21}
{Liu}, A., {Tozzi}, P., {Rosati}, P., {et~al.} 2021, arXiv e-prints,
  arXiv:2102.05788

\bibitem[{{Lodato} \& {Natarajan}(2006)}]{lodato06}
{Lodato}, G., \& {Natarajan}, P. 2006, \mnras, 371, 1813

\bibitem[{{Luo} {et~al.}(2017){Luo}, {Brandt}, {Xue}, {Lehmer}, {Alexander},
  {Bauer}, {Vito}, {Yang}, {Basu-Zych}, {Comastri}, {Gilli}, {Gu},
  {Hornschemeier}, {Koekemoer}, {Liu}, {Mainieri}, {Paolillo}, {Ranalli},
  {Rosati}, {Schneider}, {Shemmer}, {Smail}, {Sun}, {Tozzi}, {Vignali}, \&
  {Wang}}]{luo17}
{Luo}, B., {Brandt}, W.~N., {Xue}, Y.~Q., {et~al.} 2017, \apjs, 228, 2

\bibitem[{{Lusso} {et~al.}(2012){Lusso}, {Comastri}, {Simmons}, {Mignoli},
  {Zamorani}, {Vignali}, {Brusa}, {Shankar}, {Lutz}, {Trump}, {Maiolino},
  {Gilli}, {Bolzonella}, {Puccetti}, {Salvato}, {Impey}, {Civano}, {Elvis},
  {Mainieri}, {Silverman}, {Koekemoer}, {Bongiorno}, {Merloni}, {Berta}, {Le
  Floc'h}, {Magnelli}, {Pozzi}, \& {Riguccini}}]{lusso12}
{Lusso}, E., {Comastri}, A., {Simmons}, B.~D., {et~al.} 2012, \mnras, 425, 623

\bibitem[{{Madau} \& {Rees}(2001)}]{madau01}
{Madau}, P., \& {Rees}, M.~J. 2001, \apjl, 551, L27

\bibitem[{{McConnell} \& {Ma}(2013)}]{mcconnell13}
{McConnell}, N.~J., \& {Ma}, C.-P. 2013, \apj, 764, 184

\bibitem[{{Meneghetti} {et~al.}(2017){Meneghetti}, {Natarajan}, {Coe},
  {Contini}, {De Lucia}, {Giocoli}, {Acebron}, {Borgani}, {Bradac}, {Diego},
  {Hoag}, {Ishigaki}, {Johnson}, {Jullo}, {Kawamata}, {Lam}, {Limousin},
  {Liesenborgs}, {Oguri}, {Sebesta}, {Sharon}, {Williams}, \&
  {Zitrin}}]{meneghetti17}
{Meneghetti}, M., {Natarajan}, P., {Coe}, D., {et~al.} 2017, \mnras, 472, 3177

\bibitem[{{Merloni} {et~al.}(2010){Merloni}, {Bongiorno}, {Bolzonella},
  {Brusa}, {Civano}, {Comastri}, {Elvis}, {Fiore}, {Gilli}, {Hao}, {Jahnke},
  {Koekemoer}, {Lusso}, {Mainieri}, {Mignoli}, {Miyaji}, {Renzini}, {Salvato},
  {Silverman}, {Trump}, {Vignali}, {Zamorani}, {Capak}, {Lilly}, {Sanders},
  {Taniguchi}, {Bardelli}, {Carollo}, {Caputi}, {Contini}, {Coppa}, {Cucciati},
  {de la Torre}, {de Ravel}, {Franzetti}, {Garilli}, {Hasinger}, {Impey},
  {Iovino}, {Iwasawa}, {Kampczyk}, {Kneib}, {Knobel}, {Kova{\v{c}}},
  {Lamareille}, {Le Borgne}, {Le Brun}, {Le F{\`e}vre}, {Maier}, {Pello},
  {Peng}, {Perez Montero}, {Ricciardelli}, {Scodeggio}, {Tanaka}, {Tasca},
  {Tresse}, {Vergani}, \& {Zucca}}]{merloni10}
{Merloni}, A., {Bongiorno}, A., {Bolzonella}, M., {et~al.} 2010, \apj, 708, 137

\bibitem[{{Mineo} {et~al.}(2012){Mineo}, {Gilfanov}, \& {Sunyaev}}]{mineo12}
{Mineo}, S., {Gilfanov}, M., \& {Sunyaev}, R. 2012, \mnras, 419, 2095

\bibitem[{{Mortlock} {et~al.}(2011){Mortlock}, {Warren}, {Venemans}, {Patel},
  {Hewett}, {McMahon}, {Simpson}, {Theuns}, {Gonz{\'a}les-Solares}, {Adamson},
  {Dye}, {Hambly}, {Hirst}, {Irwin}, {Kuiper}, {Lawrence}, \&
  {R{\"o}ttgering}}]{mortlock11}
{Mortlock}, D.~J., {Warren}, S.~J., {Venemans}, B.~P., {et~al.} 2011, \nat,
  474, 616

\bibitem[{{Nanni} {et~al.}(2017){Nanni}, {Vignali}, {Gilli}, {Moretti}, \&
  {Brandt}}]{nanni17}
{Nanni}, R., {Vignali}, C., {Gilli}, R., {Moretti}, A., \& {Brandt}, W.~N.
  2017, \aap, 603, A128

\bibitem[{{Pons} {et~al.}(2020){Pons}, {McMahon}, {Banerji}, \&
  {Reed}}]{pons20}
{Pons}, E., {McMahon}, R.~G., {Banerji}, M., \& {Reed}, S.~L. 2020, \mnras,
  491, 3884

\bibitem[{{Postman} {et~al.}(2012){Postman}, {Coe}, {Ben{\'\i}tez}, {Bradley},
  {Broadhurst}, {Donahue}, {Ford}, {Graur}, {Graves}, {Jouvel}, {Koekemoer},
  {Lemze}, {Medezinski}, {Molino}, {Moustakas}, {Ogaz}, {Riess}, {Rodney},
  {Rosati}, {Umetsu}, {Zheng}, {Zitrin}, {Bartelmann}, {Bouwens}, {Czakon},
  {Golwala}, {Host}, {Infante}, {Jha}, {Jimenez-Teja}, {Kelson}, {Lahav},
  {Lazkoz}, {Maoz}, {McCully}, {Melchior}, {Meneghetti}, {Merten}, {Moustakas},
  {Nonino}, {Patel}, {Reg{\"o}s}, {Sayers}, {Seitz}, \& {Van der
  Wel}}]{postman12}
{Postman}, M., {Coe}, D., {Ben{\'\i}tez}, N., {et~al.} 2012, \apjs, 199, 25

\bibitem[{{Saglia} {et~al.}(2016){Saglia}, {Opitsch}, {Erwin}, {Thomas},
  {Beifiori}, {Fabricius}, {Mazzalay}, {Nowak}, {Rusli}, \&
  {Bender}}]{saglia16}
{Saglia}, R.~P., {Opitsch}, M., {Erwin}, P., {et~al.} 2016, \apj, 818, 47

\bibitem[{{Salmon} {et~al.}(2020){Salmon}, {Coe}, {Bradley}, {Bouwens},
  {Brada{\v{c}}}, {Huang}, {Oesch}, {Stark}, {Sharon}, {Trenti}, {Avila},
  {Ogaz}, {Andrade-Santos}, {Carrasco}, {Cerny}, {Dawson}, {Frye}, {Hoag},
  {Johnson}, {Jones}, {Lam}, {Lovisari}, {Mainali}, {Past}, {Paterno-Mahler},
  {Peterson}, {Riess}, {Rodney}, {Ryan}, {Sendra-Server}, {Strait}, {Strolger},
  {Umetsu}, {Vulcani}, \& {Zitrin}}]{salmon20}
{Salmon}, B., {Coe}, D., {Bradley}, L., {et~al.} 2020, \apj, 889, 189

\bibitem[{{Santini} {et~al.}(2015){Santini}, {Ferguson}, {Fontana}, {Mobasher},
  {Barro}, {Castellano}, {Finkelstein}, {Grazian}, {Hsu}, {Lee}, {Lee},
  {Pforr}, {Salvato}, {Wiklind}, {Wuyts}, {Almaini}, {Cooper}, {Galametz},
  {Weiner}, {Amorin}, {Boutsia}, {Conselice}, {Dahlen}, {Dickinson},
  {Giavalisco}, {Grogin}, {Guo}, {Hathi}, {Kocevski}, {Koekemoer},
  {Kurczynski}, {Merlin}, {Mortlock}, {Newman}, {Paris}, {Pentericci},
  {Simons}, \& {Willner}}]{santini15}
{Santini}, P., {Ferguson}, H.~C., {Fontana}, A., {et~al.} 2015, \apj, 801, 97

\bibitem[{{Schutte} {et~al.}(2019){Schutte}, {Reines}, \& {Greene}}]{schutte19}
{Schutte}, Z., {Reines}, A.~E., \& {Greene}, J.~E. 2019, \apj, 887, 245

\bibitem[{{Strait} {et~al.}(2021){Strait}, {Brada{\v{c}}}, {Coe}, {Lemaux},
  {Carnall}, {Bradley}, {Pelliccia}, {Sharon}, {Zitrin}, {Acebron}, {Neufeld},
  {Andrade-Santos}, {Avila}, {Frye}, {Mahler}, {Nonino}, {Ogaz}, {Oguri},
  {Ouchi}, {Paterno-Mahler}, {Stark}, {Mainali}, {Oesch}, {Trenti}, {Carrasco},
  {Dawson}, {Jones}, {Umetsu}, \& {Vulcani}}]{strait20}
{Strait}, V., {Brada{\v{c}}}, M., {Coe}, D., {et~al.} 2021, \apj, 910, 135

\bibitem[{{Venemans} {et~al.}(2013){Venemans}, {Findlay}, {Sutherland}, {De
  Rosa}, {McMahon}, {Simcoe}, {Gonz{\'a}lez-Solares}, {Kuijken}, \&
  {Lewis}}]{venemans13}
{Venemans}, B.~P., {Findlay}, J.~R., {Sutherland}, W.~J., {et~al.} 2013, \apj,
  779, 24

\bibitem[{{Vito} {et~al.}(2013){Vito}, {Vignali}, {Gilli}, {Comastri},
  {Iwasawa}, {Brandt}, {Alexander}, {Brusa}, {Lehmer}, {Bauer}, {Schneider},
  {Xue}, \& {Luo}}]{vito13}
{Vito}, F., {Vignali}, C., {Gilli}, R., {et~al.} 2013, MNRAS, 428, 354

\bibitem[{{Vito} {et~al.}(2016){Vito}, {Gilli}, {Vignali}, {Brandt},
  {Comastri}, {Yang}, {Lehmer}, {Luo}, {Basu-Zych}, {Bauer}, {Cappelluti},
  {Koekemoer}, {Mainieri}, {Paolillo}, {Ranalli}, {Shemmer}, {Trump}, {Wang},
  \& {Xue}}]{vito16}
{Vito}, F., {Gilli}, R., {Vignali}, C., {et~al.} 2016, \mnras, 463, 348

\bibitem[{{Vito} {et~al.}(2018){Vito}, {Brandt}, {Yang}, {Gilli}, {Luo},
  {Vignali}, {Xue}, {Comastri}, {Koekemoer}, {Lehmer}, {Liu}, {Paolillo},
  {Ranalli}, {Schneider}, {Shemmer}, {Volonteri}, \& {Wang}}]{vito18}
{Vito}, F., {Brandt}, W.~N., {Yang}, G., {et~al.} 2018, \mnras, 473, 2378

\bibitem[{{Vito} {et~al.}(2019){Vito}, {Brandt}, {Bauer}, {Calura}, {Gilli},
  {Luo}, {Shemmer}, {Vignali}, {Zamorani}, {Brusa}, {Civano}, {Comastri}, \&
  {Nanni}}]{vito19}
{Vito}, F., {Brandt}, W.~N., {Bauer}, F.~E., {et~al.} 2019, \aap, 630, A118

\bibitem[{{Volonteri} {et~al.}(2016){Volonteri}, {Dubois}, {Pichon}, \&
  {Devriendt}}]{volonteri16}
{Volonteri}, M., {Dubois}, Y., {Pichon}, C., \& {Devriendt}, J. 2016, \mnras,
  460, 2979

\bibitem[{{Volonteri} {et~al.}(2003){Volonteri}, {Haardt}, \&
  {Madau}}]{volonteri03}
{Volonteri}, M., {Haardt}, F., \& {Madau}, P. 2003, \apj, 582, 559

\bibitem[{{Volonteri} \& {Rees}(2005)}]{volonteri05}
{Volonteri}, M., \& {Rees}, M.~J. 2005, \apj, 633, 624

\bibitem[{{Wang} {et~al.}(2019){Wang}, {Wang}, {Fan}, {Wu}, {Yang}, {Neri}, \&
  {Yue}}]{wang19}
{Wang}, F., {Wang}, R., {Fan}, X., {et~al.} 2019, \apj, 880, 2

\bibitem[{{Wang} {et~al.}(2017){Wang}, {Fan}, {Yang}, {Wu}, {Yang}, {Bian},
  {McGreer}, {Li}, {Li}, {Ding}, {Dey}, {Dye}, {Findlay}, {Green}, {James},
  {Jiang}, {Lang}, {Lawrence}, {Myers}, {Ross}, {Schlegel}, \&
  {Shanks}}]{wang17}
{Wang}, F., {Fan}, X., {Yang}, J., {et~al.} 2017, \apj, 839, 27

\bibitem[{{Wang} {et~al.}(2021){Wang}, {Fan}, {Yang}, {Mazzucchelli}, {Wu},
  {Li}, {Ba{\~n}ados}, {Farina}, {Nanni}, {Ai}, {Bian}, {Davies}, {Decarli},
  {Hennawi}, {Schindler}, {Venemans}, \& {Walter}}]{wang21}
---. 2021, \apj, 908, 53

\bibitem[{{Willott} {et~al.}(2007){Willott}, {Delorme}, {Omont}, {Bergeron},
  {Delfosse}, {Forveille}, {Albert}, {Reyl{\'e}}, {Hill}, {Gully-Santiago},
  {Vinten}, {Crampton}, {Hutchings}, {Schade}, {Simard}, {Sawicki}, {Beelen},
  \& {Cox}}]{willott07}
{Willott}, C.~J., {Delorme}, P., {Omont}, A., {et~al.} 2007, \aj, 134, 2435

\bibitem[{{Wise} {et~al.}(2019){Wise}, {Regan}, {O'Shea}, {Norman}, {Downes},
  \& {Xu}}]{wise19}
{Wise}, J.~H., {Regan}, J.~A., {O'Shea}, B.~W., {et~al.} 2019, \nat, 566, 85

\bibitem[{{Yang} {et~al.}(2017){Yang}, {Fan}, {Wu}, {Wang}, {Bian}, {Yang},
  {McGreer}, {Yi}, {Jiang}, {Green}, {Yue}, {Wang}, {Li}, {Ding}, {Dye}, \&
  {Lawrence}}]{yang17}
{Yang}, J., {Fan}, X., {Wu}, X.-B., {et~al.} 2017, \aj, 153, 184

\bibitem[{{Yang} {et~al.}(2019){Yang}, {Wang}, {Fan}, {Yue}, {Wu}, {Li},
  {Bian}, {Jiang}, {Ba{\~n}ados}, \& {Beletsky}}]{yang19}
{Yang}, J., {Wang}, F., {Fan}, X., {et~al.} 2019, \aj, 157, 236

\bibitem[{{Yang} {et~al.}(2020){Yang}, {Wang}, {Fan}, {Hennawi}, {Davies},
  {Yue}, {Banados}, {Wu}, {Venemans}, {Barth}, {Bian}, {Boutsia}, {Decarli},
  {Farina}, {Green}, {Jiang}, {Li}, {Mazzucchelli}, \& {Walter}}]{yang20}
---. 2020, \apjl, 897, L14

\end{thebibliography}

\end{document}